\documentclass{kapproc} 
\usepackage{amssymb}
\usepackage{amsmath}
\usepackage{graphicx}
\kluwerbib

\def\lesssim{\mathrel{\hbox{\rlap{\hbox{\lower4pt\hbox{$\sim$}}}\hbox{$<$}}}}

\def\PLB{{\em Phys. Lett.} B}

\def\PRL{{\em Phys. Rev. Lett.}\ }

\def\PRD{{\em Phys. Rev.} D}
\def\PRC{{\em Phys. Rev.} C}

\def\apj{{\em Ap.~J.}\ }

\def\aa{{\em Astr.~Ap.}\ }
\def\aap{{\em Astr.~Ap.}\ }

\def\apjs{{\em Ap.~J.~Suppl.}\ }

\newcommand{\anu}{\bar{\nu}}
\newcommand{\aenu}{\bar{\nu}_e}
\newcommand{\enu}{\nu_e}
\newcommand{\unu}{\nu_\mu}

\newcommand{\tnu}{\nu_\tau}

\newcommand{\vep}{\epsilon}
\newcommand{\epnu}{\varepsilon_\nu}

\newcommand{\avemu}{\langle\mu_{\nu_i}\rangle}
\newcommand{\aveamu}{\langle\mu_{\bar{\nu}_i}\rangle}
\newcommand{\avemut}{\langle\mu_{\nu_i}^2\rangle}

\newcommand{\f}{{\cal{F}}}
\newcommand{\sinw}{\sin^2\theta_W}

\newcommand{\beq}{\begin{equation}} 
\newcommand{\eeq}{\end{equation}} 
\newcommand{\beqa}{\begin{eqnarray}} 
\newcommand{\eeqa}{\end{eqnarray}} 
    
\newcommand{\pr}{^\prime}    
\newcommand{\righta}{\rightarrow}
\newcommand{\sig}{\sigma}

\begin{document}

\articletitle{Neutrino Opacities in Nuclear Matter}

\author{Adam Burrows}
\affil{Steward Observatory, 
The University of Arizona, Tucson, AZ 85721 \\}
\email{burrows@zenith.as.arizona.edu}

\author{Sanjay Reddy}
\affil{Theoretical Division, Los Alamos National Laboratory, Los Alamos, NM 87544\\}
\email{reddy@lanl.gov}

\author{Todd A. Thompson}
\affil{Astronomy Department and Theoretical Astrophysics Center,
                 The University of California, Berkeley, CA 94720\\}
\email{thomp@astro.berkeley.edu}

\begin{abstract}
Neutrino-matter cross sections and interaction rates are 
central to the core-collapse supernova phenomenon and, very 
likely, to the viability of the explosion mechanism itself.  In
this paper, we describe the major neutrino scattering,
absorption, and production processes that together influence
the outcome of core collapse and the cooling of protoneutron
stars.  One focus is on energy redistribution and many-body physics,
but our major goal is to provide a useful resource for those 
interested in supernova neutrino microphysics.
\end{abstract}

\begin{keywords}
Supernovae, Neutrino Interactions, Neutrino Spectra, Protoneutron Stars, Kinetic Theory
\end{keywords}

\section{Introduction}

Supernova explosions are one major means by which elements
are injected into the interstellar medium and, hence, 
into subsequent generations of stars.  Therefore, supernovae are
central to the chemical evolution and progressive enrichment of the universe. 
Most supernova explosions are the outcome of the dynamical collapse of
the core of a massive star as it dies.  Collapse creates high temperatures ($> 1$ MeV)
and densities ($10^7$ g cm$^{-3} < \rho < 10^{15}$ g cm$^{-3}$) and produces
(``after the dust settles") either a neutron star or a black hole.  Under
such extreme thermodynamic conditions, neutrinos are produced in abundance.
The mechanism of core-collapse supernovae is thought to depend 
upon the transfer of energy from the inner core to the
outer mantle of the iron core of the massive star.  Neutrinos seem to
be the mediators of this energy transfer.
Therefore, to fully understand core-collapse supernova explosions one must have a firm handle on 
the physics of neutrino production, absorption, and scattering.  

In this paper, we summarize the neutrino-matter cross sections 
and the neutrino production rates in the core-collapse context.
Some of this discussion can already be found in Burrows (2001).
We do not attempt to explain the hydrodynamics of supernova explosions, but do try
to present the relevant neutrino processes that play a role.  For the former, the reader is
referred to \cite{nature}, \cite{bhf_1995}, \cite{lieben2001}, 
\cite{lieben20012}, \cite{rampp20022}, \cite{buras2003}, and Thompson, Burrows, \& Pinto (2003).

In \S\ref{stimabs}, we present   
a physical derivation of stimulated absorption and then in \S\ref{cross6} 
we summarize the basic neutrino-matter cross sections.  In \S\ref{section:inelastic}, we discuss
the neutrino-electron scattering kernel, along with a simple treatment of the collision
integral.   In \S\ref{freegas}, we provide the
relativisitic formalism for inelastic neutrino-electron and neutrino-nucleon scattering 
processes and energy redistribution for 
non-interacting nucleons.  This is followed in \S\ref{strongandeandm} with 
a discussion of the alternate, more powerful, formalism 
for determining differential interaction rates and 
redistribution in the many-body context, namely that of 
dynamical structure factors.  The role of strong and 
electromagnetic interactions between nucleons and leptons
is explored, as well as collective excitations of the medium.
Source terms for electron-positron annihilation (\S\ref{eplus}), 
neutrino-anti-neutrino annihilation (\S\ref{paira}),
and nucleon-nucleon bremsstrahlung (\S\ref{bremsst}) cap off our review of the major
processes of relevance in core-collapse simulations.

\section{Stimulated Absorption}
\label{stimabs}

The concept of stimulated emission for photons is well understood and studied, but
the corresponding concept of stimulated {\it absorption} for neutrinos is not so
well appreciated.  This may be because its simple origin in Fermi blocking and the
Pauli exclusion principle in the context of {\it net} emission is not often explained.
The {\it net} emission of a neutrino is simply the difference between the emissivity and
the absorption of the medium:
\beq
{\cal {J}}_{net}=\eta_\nu-\kappa_aI_\nu\,\, ,
\label{netemission}
\eeq
where $\kappa_a$ is the absorptive opacity and $I_\nu$ is the specific intensity.
All absorption processes involving fermions will be inhibited by Pauli blocking due to
final--state occupancy.  Hence, $\eta_\nu$ in eq. (\ref{netemission}) includes a blocking term,
$(1-\f_{\nu})$(Bruenn 1985). $\f_{\nu}$ is the invariant distribution function
for the neutrino, whether or not it is in chemical equilibrium.

We can derive stimulated absorption using Fermi's Golden rule.  For example, the net collision term
for the process, $\nu_en\leftrightarrow e^-p$, is:
\beqa
{\cal C}_{\nu_e n\leftrightarrow e^- p}&=&
\int\frac{d^3\vec{p}_{\nu_e}}{(2\pi)^3 }
\int\frac{d^3\vec{p}_n}{(2\pi)^3 }
\int\frac{d^3\vec{p}_p}{(2\pi)^3 }
\int\frac{d^3\vec{p}_e}{(2\pi)^3 }\,
\left(\sum_{s}{|{\cal{M}}|}^2\right) \nonumber \\
&& \nonumber \\
&\times&
\Xi(\nu_en\leftrightarrow e^-p)\,\,(2\pi)^4\,\delta^4({\bf p}_{\nu_e}+{\bf
p}_n-{\bf p}_p-{\bf p}_e)
\,\, ,
\label{rate01}
\eeqa
where ${\bf p}$ is a four-vector and
\beq
\Xi(\nu_en\leftrightarrow
e^-p)=\f_{\nu_e}\f_n(1-\f_e)(1-\f_p)-\f_e\f_p(1-\f_n)(1-\f_{\nu_e})\,\, .
\label{blocks}
\eeq
The final--state blocking terms in eq. (\ref{blocks}) are manifest, in particular that for the $\nu_e$ neutrino.
Algebraic manipulations convert $\Xi(\nu_en\leftrightarrow e^-p)$ in eq. (\ref{blocks}) into:
\beqa
\Xi(\nu_en\leftrightarrow e^-p)&=&
\f_n(1-\f_e)(1-\f_p)\left[\frac{\f_{\nu_e}^{\rm eq}}{1-\f_{\nu_e}^{\rm eq}}(1-\f_{\nu_e})-\f_{\nu_e}\right]
\nonumber \\
&=&\frac{\f_n(1-\f_e)(1-\f_p)}{1-\f_{\nu_e}\pr}\left[\f_{\nu_e}^{\rm eq}-\f_{\nu_e}\right]\,\, ,
\label{kirch03}
\eeqa
where
\beq
\f_{\nu_e}^{\rm eq}=[e^{(\varepsilon_{\nu_e}-(\mu_e-\hat{\mu}))\beta}+1]^{-1}
\label{equileq}
\eeq
is an equilibrium distribution function for the $\nu_e$ neutrino
and it has been assumed that only the electron, proton, and neutron
are in thermal equilibrium.  $\hat{\mu}$ is the difference between the neutron
and the proton chemical potentials.  Note that in $\f_{\nu_e}^{\rm eq}$
there is no explicit reference to a neutrino chemical potential, though
of course in beta equilibrium it is equal to $\mu_e-\hat{\mu}$.
There is no need to construct or
refer to a neutrino chemical potential in neutrino transfer.

We see that eq. (\ref{kirch03}) naturally leads to:
\beq
{\cal {J}}_{net}=\frac{\kappa_a}{1-\f^{\rm eq}_\nu}\left(B_\nu-I_\nu\right)=\kappa_a^*(B_\nu-I_\nu)\,\, .
\label{kirch}
\eeq
Of course, $B_\nu$ is the black body function for neutrinos.
This expression emphasizes the fact that ${\cal C}_{\nu_e n\leftrightarrow e^- p}$ and ${\cal {J}}_{net}$ are the same entity.
If neutrinos were bosons, we would have found a (${1+\f^{\rm eq}_\nu}$) in the denominator, but the
form of eq. (\ref{kirch}) in which $I_\nu$ is manifestly driven to $B_\nu$, the equilibrium
intensity, would have been retained.  From eqs. (\ref{kirch03}) and (\ref{kirch}), we
see that the stimulated absorption correction to $\kappa_a$ is $1/(1-\f^{\rm eq}_\nu)$.
By writing the collision term in the form of eq. (\ref{kirch}), with $\kappa_a$ corrected for stimulated
absorption, we have a net source term that clearly drives $I_\nu$ to equilibrium.  The timescale
is $1/c\kappa_a^*\,$.   Though the derivation of the stimulated absorption correction
we have provided here is for the $\nu_en\leftrightarrow e^-p$ process, this correction is quite general and
applies to all neutrino absorption opacities.

Kirchhoff's Law, expressing detailed balance, is:
\beq
\kappa_a = \eta_\nu/B_\nu \,\,{\rm or}\,\, \kappa^*_a = \eta^{\prime}_\nu/B_\nu\, ,
\label{kirchh}
\eeq
where $\eta^{\prime}_\nu$ is not corrected for final--state neutrino blocking.
Furthermore, the net emissivity can be written as the sum of its {\it spontaneous} and {\it
induced} components:
\beq
\eta_\nu=\kappa_a\left[\frac{B_\nu}{1\pm\f^{\rm eq}_\nu}+\left(1-\frac{1}{1\pm\f^{\rm eq}_\nu}\right)I_\nu\right]\,\,  ,
\label{emission}
\eeq
where $+$ or $-$ is used for bosons or fermions, respectively.
Eq.(\ref{kirchh}) can be used to convert the absorption cross sections described in \S\ref{cross6}
into source terms.

\section{Neutrino Cross Sections}
\label{cross6}

Neutrino--matter cross sections, both for scattering and for absorption,
play the central role in neutrino transport.  The major processes are
the super--allowed charged--current absorptions of $\nu_e$ and $\bar{\nu}_e$
neutrinos on free nucleons, neutral--current scattering off of free nucleons,
alpha particles, and nuclei (Freedman 1974), neutrino--electron/positron scattering,
neutrino--nucleus absorption, neutrino--neutrino scattering, neutrino--antineutrino absorption, and the inverses
of various neutrino production processes such as nucleon--nucleon bremsstrahlung
and the modified URCA process ($\nu_e + n + n \rightarrow e^- + p + n$).
Compared with photon--matter interactions, neutrino--matter interactions
are relatively simple functions of incident neutrino energy.
Resonances play little role and continuum processes dominate. Nice early summaries
of the various neutrino cross sections of relevance in supernova
theory are given in \cite{tubbs_schramm} and in \cite{bruenn_1985}.
In particular, \cite{bruenn_1985} discusses in detail
neutrino--electron scattering and neutrino--antineutrino processes
using the full energy redistribution formalism.  He also provides a practical approximation
to the neutrino--nucleus absorption cross section (Fuller, Fowler, \& Newman 1982; 
Aufderheide et al.~1994), though better physics soon promises to improve 
the treatment of these cross sections
substantially (Langanke \& Martinez-Pinedo 2003; Pruet \& Fuller 2003). 
For a neutrino energy of $\sim$10 MeV the ratio of the charged--current cross section
to the $\nu_e$--electron scattering cross section is $\sim$100.
However, neutrino--electron scattering does play a role, along with neutrino--nucleon
scattering and nucleon--nucleon bremsstrahlung, in the energy 
equilibration of emergent $\nu_\mu$ neutrinos (Thompson, Burrows, \& Horvath 2000). 

Below, we list and discuss many of the absorption and elastic scattering 
cross sections one needs in detailed supernova calculations.
In \S\ref{section:inelastic}, \S\ref{freegas}, and \S\ref{strongandeandm}, we provide some straightforward
formulae that can be used to properly handle inelastic scattering.
The set of these processes comprises the essential microphysical 
package for the simulation of neutrino atmospheres and core--collapse supernovae.

\subsection{{\bf $\enu\,+\,n\, \righta\, e^-\,+\,p$:}}
\label{CCabs}
The cross section per baryon for $\nu_e$ neutrino absorption on free neutrons is larger than that
for any other process.   Given the large abundance of free neutrons in protoneutron star atmospheres,
this process is central to $\nu_e$ neutrino transport.
A convenient reference neutrino cross section is $\sigma_o$, given by
\beq
\sigma_o\,=\,\frac{4G_F^2(m_ec^2)^2}{\pi(\hbar c)^4}\simeq \,1.705\times
10^{-44}\,cm^2\,\, ,
\eeq
where $G_F$ is the Fermi weak coupling constant ($\simeq 1.436\times 10^{-49}$ ergs cm$^{-3}$).
The total $\enu- n$ absorption cross section is then given by
\beq
\sigma^a_{\enu
n}=\,\sig_o\left(\frac{1+3g_A^2}{4}\right)\,\left(\frac{\varepsilon_{\nu_e}+\Delta_{np}}{m_ec^2}\right)^2\,
\Bigl[1-\left(\frac{m_ec^2}{\varepsilon_{\nu_e}+\Delta_{np}}\right)^2\Bigr]^{1/2}W_M\,\, ,
\label{ncapture}
\eeq
where $g_A$ is the axial--vector coupling constant ($\sim -1.23$), $\Delta_{np}=m_nc^2-m_pc^2=1.29332$
MeV, and for a collision in which the electron gets all of the
kinetic energy $\vep_{e^-}=\varepsilon_{\nu_e}+\Delta_{np}$.
$W_M$ is the correction for weak magnetism and recoil (Vogel 1984; \S\ref{freegas}) 
and is approximately equal to $(1 + 1.1\varepsilon_{\nu_e}/m_nc^2)$.
At $\varepsilon_{\nu_e} = 20$ MeV, this correction is only $\sim2.5$\%.  We include it here for
symmetry's sake, since the corresponding correction ($W_{\bar{M}}$) for $\bar{\nu}_e$ neutrino absorption
on protons is $(1 - 7.1\varepsilon_{\bar{\nu}_e}/m_nc^2)$, which at 20 MeV is a large $-15$\%.
To calculate $\kappa_a^*$, $\sigma^a_{\enu n}$ must be multiplied by the stimulated absorption
correction, $1/(1-\f_{\nu_e}\pr)$, and final--state blocking
by the electrons and the protons \`a la eq. (\ref{kirch03}) must be included.

\subsection{{\bf $\aenu\,+\,p\, \righta\, e^+\,+\,n$:}}

The total $\aenu- p$ absorption cross section is given by
\beq
\sigma^a_{\aenu
p}=\sig_o\left(\frac{1+3g_A^2}{4}\right)\,\left(\frac{\vep_{\bar{\nu}_e}-\Delta_{np}}{m_ec^2}\right)^2\,
\Bigl[1-\left(\frac{m_ec^2}{\vep_{\bar{\nu}_e}-\Delta_{np}}\right)^2\Bigr]^{1/2}W_{\bar{M}}\, ,
\label{pcapture}
\eeq
where $\vep_{e^+}=\vep_{\bar{\nu_e}}-\Delta_{np}$ and $W_{\bar{M}}$ is the weak
magnetism/recoil correction given in \S\ref{CCabs}.  Note that $W_{\bar{M}}$ 
is as large as many other corrections and should not be ignored.  To calculate
$\kappa_a^*$, $\sigma^a_{\aenu p}$ must also be corrected for stimulated absorption and final--state blocking.
However, the sign of $\mu_e -\hat{\mu}$ in the stimulated absorption correction for $\bar{\nu}_e$ neutrinos
is flipped, as is the sign of $\mu_e$ in the positron blocking term.  Hence, as a consequence of the severe electron
lepton asymmetry in core--collapse supernovae, both coefficients are very close to one.
Note that the $\aenu\,+\,p\, \righta\, e^+\,+\,n$ process
dominates the supernova neutrino signal in proton--rich underground neutrino
telescopes on Earth, such as Super Kamiokande, SNO, and LVD, a fact that emphasizes the
interesting complementarities between emission at the supernova and detection in \v Cerenkov and scintillation
facilities.

\subsection{$\enu A \leftrightarrow A\pr e^-$}

From \cite{bruenn_1985} the total $\nu_e - A$ absorption cross section,
is approximated by
\begin{equation}
\sigma^a_A=\frac{\sigma_o}{14}\,
g_A^2\,N_p(Z)\,N_n(N)\,\left(\frac{\varepsilon_\nu+Q\pr}{m_ec^2}\right)^2\,
\left[1-\left(\frac{m_e c^2}{\varepsilon_\nu+Q\pr}\right)^2\right]^{1/2}
W_{block}
\,\, ,
\label{block}
\end{equation}
where $W_{block} = \left(1-f_{e^-}\right)\,e^{(\mu_n-\mu_p-Q\pr)\beta}$,
$Q\pr=M_{A\pr}-M_A+\Delta\sim\mu_n-\mu_p+\Delta$, $\Delta$ is
the energy of the neutron $1f_{5/2}$ state above the ground state
and is taken to be 3 MeV (Fuller 1982), and
the quantities $N_p(Z)$ and $N_n(N)$ are approximated by:
$N_p(Z)=0$, $Z-20$, and 8 for $Z<20$, $20<Z<28$, and $Z>28$,
respectively, and $N_n(N)=6$, $40-N$, and 0 for $N<34$, $43<N<40$, and $N>40$, respectively.
The opacity, corrected for stimulated absorption, is then
\begin{equation}
\kappa_a^*=X_H \,\rho N_A\sigma^a_A(1-\f_{\nu_e}^{\rm eq})^{-1}.
\end{equation}
Since $N_n(N)=0$ for $N>40$, this absorption and emission
process plays a role only during the very early phase of collapse.
Typically, at densities near $\rho\sim10^{12}$ g cm$^{-3}$, $\kappa_a^*\rightarrow 0$.
Eq. \ref{block} is only approximate and better estimates 
of the $\nu_e - A$ absorption cross section are in the 
offing (Langanke \& Martinez-Pinedo 2003; Pruet \& Fuller 2003).

\subsection{{\bf $\nu_i\,+\,p\,\righta\,\nu_i\,+\,p$:}} 
\label{nupro}

The total $\nu_i - p$ elastic scattering cross section for all neutrino species is:
\beq
\sigma_{p}=\frac{\sigma_o}{4}\left(\frac{\epnu}{m_ec^2}\right)^2
\left(4\sin^4\theta_W-2\sinw+\frac{(1+3g_A^2)}{4}\right)\,\, ,
\label{pscatter}
\eeq
where $\theta_W$ is the Weinberg angle and $\sinw \simeq 0.23$.
In terms of $C_V\pr=1/2+2\sinw$ and $C_A\pr=1/2$ (note primes), eq. (\ref{pscatter}) becomes (Schinder 1990):
\beq
\sigma_{p}=\frac{\sigma_o}{4}\left(\frac{\epnu}{m_ec^2}\right)^2
\left[(C_V\pr-1)^2+3g_A^2(C_A\pr-1)^2\right].
\eeq
The differential cross section is:
\beq
\frac{d\sig_p}{d\Omega}=\frac{\sigma_{p}}{4\pi}(1+\delta_p\mu)\,\, ,
\label{nupscatt}
\eeq
where
\beq
\delta_p=\frac{(C_V\pr-1)^2-g_A^2(C_A\pr-1)^2}{(C_V\pr-1)^2+3g_A^2(C_A\pr-1)^2}\,\, .
\eeq
Note that $\delta_p$, and $\delta_n$ below, are negative ($\delta_p\sim-0.2$ and $\delta_n\sim-0.1$)
and, hence, that these processes are backward--peaked.

The transport (or momentum-transfer) cross section is simply
\beq
\sig^{tr}_p=\frac{\sigma_o}{6}\left(\frac{\epnu}{m_ec^2}\right)^2
\left[(C_V\pr-1)^2+5g_A^2(C_A\pr-1)^2\right]\,\, .
\eeq
where
\begin{equation}
\sigma^{tr}_i=\int\frac{d\sigma_i}{d\Omega}(1-\mu)\,d\Omega
=\sigma_i\left(1-\frac{1}{3}\delta_i\right)\,\,.
\label{transcs}
\end{equation}

\subsection{{\bf $\nu_i\,+\,n\,\righta\,\nu_i\,+\,n$:}} 
\label{nuneu}

The total $\nu_i - n$ elastic scattering cross section is:
\beq
\sigma_{n}=\frac{\sig_o}{4}\left(\frac{\epnu}{m_ec^2}\right)^2
\left(\frac{1+3g_A^2}{4}\right)\,\, .
\label{nunscatt}
\eeq
The corresponding differential cross section is:
\beq
\frac{d\sig_n}{d\Omega}=\frac{\sigma_{n}}{4\pi}(1+\delta_n\mu)\,\, ,
\label{nundiffscatt}
\eeq
where
\beq
\delta_n=\frac{1-g_A^2}{1+3g_A^2}\,\, .
\eeq
The transport cross section is
\beq
\sig^{tr}_n=\frac{\sigma_o}{4}\left(\frac{\epnu}{m_ec^2}\right)^2
\left(\frac{1+5g_A^2}{6}\right)\,\, .
\label{nuntransscatt}
\eeq
The fact that $\delta_p$ and $\delta_n$ are negative and, as a consequence, that
$\sig^{tr}_i$ is greater than $\sigma_i$ increases the neutrino--matter energy coupling rate
for a given neutrino flux in the semi--transparent region.  

\cite{horowitz02} derived expressions that include a weak magnetism/recoil
correction analogous to those previously discussed for the charged-current absorption
rates $\nu_e n\leftrightarrow p e^-$ and $\bar{\nu}_e n\leftrightarrow n e^+$.
We take the following form for the weak magnetism/recoil correction,
a fit to the actual correction factor for the transport cross sections:
\begin{equation}
\sigma^{tr}_{n,p}\rightarrow\sigma^{tr}_{n,p}(1+C_{W_M}\varepsilon_\nu/m_{n,p}),
\end{equation}
where for neutrino-neutron scattering $C_{W_M}\simeq-0.766$, for neutrino-proton scattering
$C_{W_M}\simeq-1.524$, for anti-neutrino-neutron scattering $C_{W_M}\simeq-7.3656$, and
for anti-neutrino-proton scattering $C_{W_M}\simeq-6.874$.

In fact, neutrino-nucleon scattering is slightly inelastic and when this is germane,
as with $\unu$ and $\tnu$ neutrinos, the more general formalism of \S\ref{section:inelastic}, \S\ref{freegas},
\S\ref{strongandeandm} is necessary.

\subsection{{\bf $\nu_i\,+\,A\,\righta\,\nu_i\,+\,A$:}}
\label{nuA}
In the post--bounce phase, nuclei exist in the unshocked region exterior to the shock.
At the high entropies in shocked protoneutron star atmospheres
there are very few nuclei.  There are alpha particles, but their fractional abundances are generally low, growing to
interesting levels due to reassociation of free nucleons just interior to the shock only at late times.
However, nuclei predominate on infall and neutrino-nucleus scattering (Freedman 1974)
is the most important process during the lepton trapping phase.

The differential $\nu_i - A$ neutral--current scattering cross section may be expressed as:
\beq
\frac{d\sig_{A}}{d\Omega}=\frac{\sig_o}{64\pi}
\left(\frac{\varepsilon_\nu}{m_ec^2}\right)^2\,A^2\,
\left\{{\cal W}\,{\cal C}_{FF}+{\cal
C}_{LOS}\right\}^2\,\langle {\cal S}_{ion}\rangle\,(1+\mu)\,\, ,
\label{nunucleuscs}
\eeq
where
\beq
{\cal W} = 1-\frac{2Z}{A}(1-2\sinw) \, ,
\eeq
$Z$ is the atomic number, $A$ is the atomic weight, and $\langle {\cal S}_{ion}\rangle$ is the
ion--ion correlation function, determined mostly by the Coulomb interaction between the
nuclei during infall.
$\langle {\cal S}_{ion}\rangle$,
in eq. (\ref{nunucleuscs}) was investigated by
\cite{horowitz97} who approximated it with the expansion
\begin{equation}
\langle{\cal{S}}_{ion}(\epsilon)\rangle=
\left[1+exp\left(-\sum_{i=0}^6\beta_i(\Gamma)\epsilon^i\right)\right]^{-1}\,\, ,
\end{equation}
where
\begin{equation}
\Gamma=\frac{(Ze)^2}{a}\frac{1}{kT} \hspace{.75cm},\hspace{.75cm}
\epsilon_i=\frac{\varepsilon_{\nu_i}}{\hbar c a}\hspace{.75cm},\hspace{.75cm}
a=\left(\frac{3}{4\pi n_{ion}}\right)^{1/3},
\end{equation}
$a$ is the interparticle spacing, $n_{ion}$ is the number density of ions, $\Gamma$
is the ratio of the Coulomb potential between ions to the thermal energy in the medium,
and $\beta_i$ are specified functions of $\Gamma$ for each neutrino species.

\cite{los88} have investigated the electron polarization correction, ${\cal C}_{LOS}$, and
find that
\beq
{\cal C}_{LOS}=
\frac{Z}{A}\left(\frac{1+4\sin^2\theta_W}{1+(kr_D)^2}\right)\,\, ,
\eeq
where the Debye radius is
\beq
r_D=\sqrt{\frac{\pi\hbar^2 c}{4\alpha p_F E_F}} \,\, ,
\eeq
$k^2=|{\bf p-p\pr}|^2=2(\varepsilon_\nu/c)^2(1-\mu)$, $p_F$ and $E_F$ are the electron Fermi momentum
and energy,  and $\alpha$ is the fine--structure constant ($\simeq 137^{-1}$).  Note that
$r_D\sim 10\hbar/p_F$ in the ultra--relativistic limit ($p_F>>m_e c$).
The ${\cal C}_{LOS}$ term is important only for low neutrino energies, generally below $\sim5$ MeV.

Following \cite{tubbs_schramm} and \cite{bml81}, the form factor term, ${\cal
C}_{FF}$, in eq. (\ref{nunucleuscs}) can be approximated by:
\beq
{\cal C}_{FF}=e^{-y(1-\mu)/2}\,\, ,
\eeq
where
$$y=\frac{2}{3}\varepsilon_\nu^2\langle r^2 \rangle/(\hbar c)^2\simeq\left(\frac{\epnu}{56\,
{\rm{MeV}}}\right)^2\left(\frac{A}{100}\right)^{2/3}\, , $$
and $\langle r^2 \rangle^{1/2}$ is the {\it rms} radius of the nucleus.
${\cal C}_{FF}$ differs from $1$ for large $A$ and $\varepsilon_\nu$, when the de Broglie
wavelength of the neutrino is smaller than the nuclear radius.

When $\langle {\cal S}_{ion}\rangle={\cal C}_{FF}={\cal C}_{LOS}+1=1$,
we have simple coherent Freedman scattering.   The physics of the polarization,
ion--ion correlation, and form factor corrections to coherent scattering
is interesting in its own right, but has little effect on supernovae (Bruenn \& Mezzacappa 1997).
The total and transport scattering cross sections for $\nu_i - \alpha$ scattering ($Z=2$;$A=4$) are simply
\begin{equation}
\sigma_\alpha=\frac{3}{2}\sigma^{tr}_\alpha=4\,\sigma_o\left(\frac{\epnu}{m_ec^2}\right)^2\sin^4\theta_W.
\end{equation}

\section{The Neutrino Scattering Kernel and a Simplified Approach to Inelastic Scattering}
\label{section:inelastic}

Many authors have studied inelastic neutrino-electron scattering as
an important energy redistribution process which helps to thermalize neutrinos and
increase their energetic coupling to matter in supernova explosions 
(Bruenn 1985; Mezzacappa \& Bruenn 1993abc).
Comparatively little attention has been paid to inelastic
neutrino-nucleon scattering.  \cite{thompson} and \cite{raffelt}
showed that, at least for $\unu$ and $\tnu$ neutrinos, this process cannot 
be ignored.  Here, we review the Legendre expansion formalism
for approximating the angular dependence of the scattering kernel, detail our own
implementation of scattering terms in the Boltzmann equation, and
include a discussion of neutrino-nucleon energy
redistribution. In \S\ref{freegas}, we present an alternate approach 
involving dynamical structure factors that is more easily generalized
to include many-body effects (\S\ref{strongandeandm}).

The general collision integral (RHS of the Boltzmann equation) for inelastic scattering may be written as
\begin{equation}
{\cal L}^{\rm scatt}_\nu[f_\nu]=(1-f_\nu)\int\frac{d^3p_\nu\pr}{c(2\pi\hbar c)^3}f_\nu\pr\,
R^{\rm in}(\varepsilon_\nu,\varepsilon_\nu\pr,\cos\theta) \nonumber
\end{equation}
\begin{equation}
\hspace*{4.8cm}-f_\nu\int\frac{d^3p_\nu\pr}{c(2\pi\hbar c)^3}(1-f_\nu\pr)\,
R^{\rm out}(\varepsilon_\nu,\varepsilon_\nu\pr,\cos\theta)
\label{gencoll}
\end{equation}
\begin{equation}
=\tilde{\eta}_\nu^{\rm scatt}-\tilde{\chi}_\nu^{\rm scatt}f_\nu\hspace{2.5cm}
\label{etatilde}
\end{equation}
where $\cos\theta$ is the cosine of the scattering angle, $\varepsilon_\nu$ is the
incident neutrino energy,
and $\varepsilon_\nu\pr$ is the scattered neutrino energy.   Although we suppress it here, the incident
and scattered neutrino phase space distribution functions ($f_\nu$ and $f_\nu\pr$, respectively)
have the following dependencies: $f_\nu=f_\nu(r,t,\mu,\varepsilon_\nu)$ and
$f_\nu\pr=f_\nu\pr(r,t,\mu\pr,\varepsilon_\nu\pr)$.  
$\mu$ and $\mu\pr$ are the cosines of the angular
coordinate of the zenith angle in spherical symmetry and are related to $\cos\theta$ through
\begin{equation}
\cos\theta=\mu\mu\pr+[(1-\mu^2)(1-\mu^{\prime\,2})]^{1/2}\cos(\phi-\phi\pr).
\label{costheta}
\end{equation}
The only difference
between $f_\nu$ and $\f_{\nu}$ in \S\ref{stimabs} is that here $f_\nu$ has explicit $\mu$
and $\varepsilon_\nu$ dependencies.
$R^{\rm in}$ is the scattering kernel for scattering into the
bin ($\varepsilon_\nu$, $\mu$) from any bin ($\varepsilon_\nu\pr$, $\mu\pr$)
and $R^{\rm out}$ is the scattering kernel for scattering out of the
bin ($\varepsilon_\nu$, $\mu$) to any bin ($\varepsilon_\nu\pr$, $\mu\pr$).
The kernels are Green's functions that connect points
in energy and momentum space.  One may also write
$R(\varepsilon_\nu,\varepsilon_\nu\pr,\cos\theta)$ as $R(q,\omega)$, where
$\omega(=\varepsilon_\nu-\varepsilon_\nu\pr)$ is the energy transfer and
$q(=[\varepsilon_\nu^2+\varepsilon_\nu^{\prime\,2}-2\varepsilon_\nu\varepsilon_\nu\pr\cos\theta]^{1/2})$
is the momentum transfer, so that the kernel explicitly reflects these dependencies (\S\ref{freegas}).

\begin{figure}
\begin{center}
\resizebox{.48\textwidth}{!}{%
\includegraphics{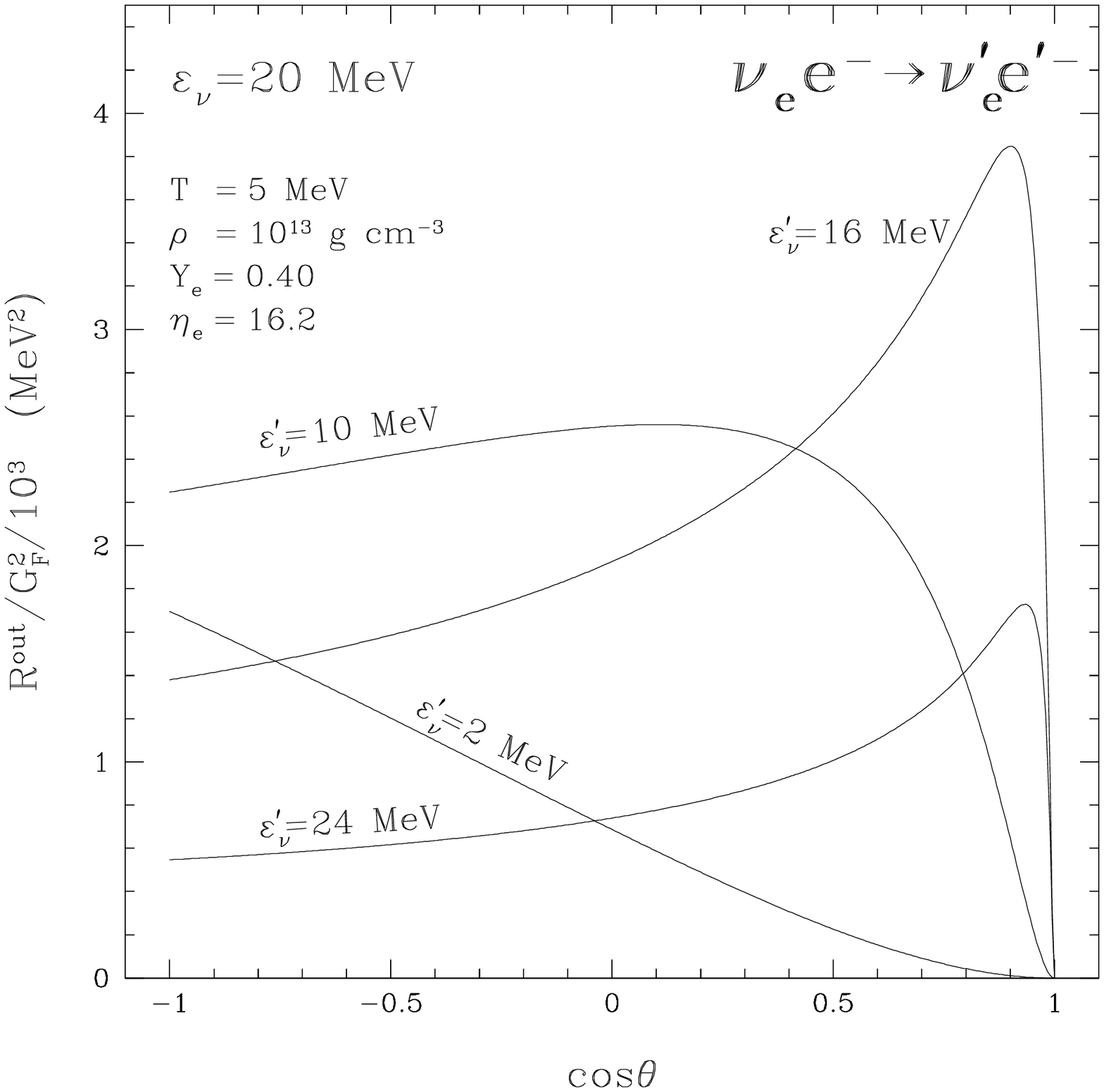}
}
\caption[$R^{\rm out}(\varepsilon_\nu,\varepsilon_\nu\pr,\cos\theta)$ for $\nu_e e^-$
scattering vs.~$\cos\theta$]
{The scattering kernel 
$R^{\rm out}(\varepsilon_\nu,\varepsilon_\nu\pr,\cos\theta)$ for
$\nu_e-$electron scattering as a function of
$\cos\theta$ for $\varepsilon_\nu=20$ MeV and $\varepsilon_\nu\pr=2$, 10, 16, and 24 MeV,
at a representative thermodynamic point ($T=5$ MeV, $\rho=10^{13}$ g cm$^{-3}$, $Y_e=0.4$).
}
\label{ek1}
\end{center}
\end{figure}

\begin{figure}
\begin{center}
\resizebox{.48\textwidth}{!}{%
\includegraphics{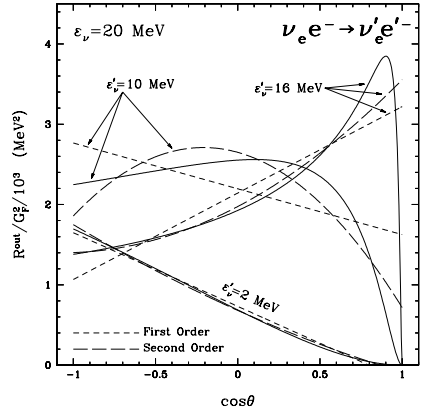}
}
\caption[$R^{\rm out}(\varepsilon_\nu,\varepsilon_\nu\pr,\cos\theta)$ for $\nu_e e^-$
scattering vs.~$\cos\theta$ with 0th, 1st, \& 2nd-order Legendre expansions]
{For the same thermodynamic point as used for Fig. \ref{ek1},
the scattering kernel ($R^{\rm out}$, thick solid lines) for $\nu_e-$electron scattering 
as a function of $\cos\theta$, 
for $\varepsilon_\nu=20$ MeV and $\varepsilon_\nu\pr=2$, 10, and 16 MeV. 
Short dashed lines show the first-order Legendre series expansion approximation
to $R^{\rm out}$, which is linear in $\cos\theta$; $R^{\rm out}\sim(1/2)\Phi_0+(3/2)\Phi_1\cos\theta$.
The long dashed line shows the improvement in going to second order in $\cos\theta$
by taking $R^{\rm out}\sim(1/2)\Phi_0+(3/2)\Phi_1\cos\theta+
(5/2)\Phi_2(1/2)(3\cos^2\theta-1)$.
}
\label{ek2}
\end{center}
\end{figure}

\begin{figure}
\begin{center}
\resizebox{.48\textwidth}{!}{%
\includegraphics{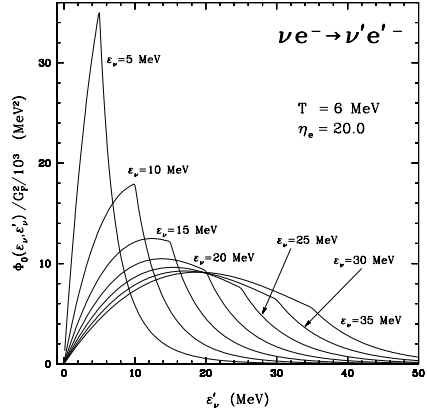}
}
\caption[$\Phi_0(\varepsilon_\nu,\varepsilon_\nu\pr)$ vs.~$\varepsilon_\nu\pr$ 
for $\nu_e e^-$ scattering for various $\varepsilon_\nu$]
{The $l=0$ term in the Legendre expansion of the $\nu_e-$electron scattering
kernel, $\Phi_0(\varepsilon_\nu,\varepsilon_\nu\pr)$ (eq. (\ref{momentkernel})), 
for $T=6$ MeV and $\eta_e=20$ as a 
function of $\varepsilon_\nu\pr$ for $\varepsilon_\nu=5$, 10, 15, 20, 25, and 35 MeV.
Note that for any $\varepsilon_\nu$, the neutrino is predominantly downscattered.
The magnitude of $\Phi_0(\varepsilon_\nu,\varepsilon_\nu\pr)$ and sign of $\langle\omega\rangle$
are to be compared with those in Fig. \ref{nk3} for $\nu$-nucleon scattering.}
\label{ek3}
\end{center}
\end{figure}

\begin{figure}
\begin{center}
\resizebox{.48\textwidth}{!}{%
\includegraphics{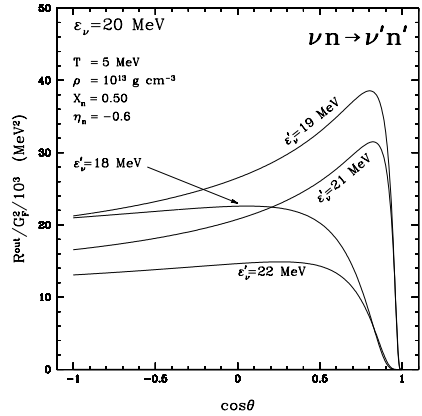}
}
\caption[$R^{\rm out}(\varepsilon_\nu,\varepsilon_\nu\pr,\cos\theta)$ for $\nu_e n$
scattering vs.~$\cos\theta$]
{The scattering kernel 
$R^{\rm out}(\varepsilon_\nu,\varepsilon_\nu\pr,\cos\theta)$ for
$\nu_e-$neutron scattering as a function of
$\cos\theta$ for $\varepsilon_\nu=20$ MeV and $\varepsilon_\nu\pr=18$, 19, 21, and 22 MeV,
at a representative thermodynamic point ($T=5$ MeV, $\rho=10^{13}$ g cm$^{-3}$, $X_n=0.5$).
Note that although the absolute value of the energy transfer 
($|\varepsilon_\nu-\varepsilon_\nu\pr|$) is the same for 
both $\varepsilon_\nu\pr=19$ MeV and 
$\varepsilon_\nu\pr=21$, the absolute value of $R^{\rm out}(20,19,\cos\theta)$ is
greater than that of $R^{\rm out}(20,21,\cos\theta)$, reflecting the fact that
at this temperature the incoming neutrino is more likely to downscatter than upscatter.
}
\label{nk1}
\end{center}
\end{figure}

\begin{figure}
\begin{center}
\resizebox{.48\textwidth}{!}{%
\includegraphics{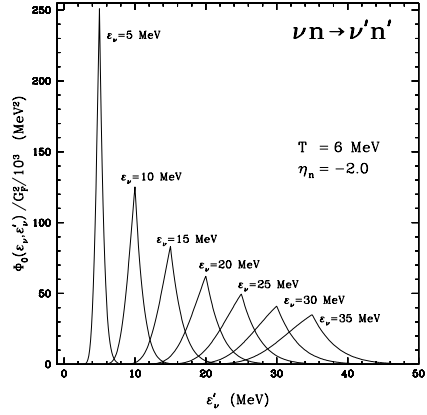}
}
\caption[$\Phi_0(\varepsilon_\nu,\varepsilon_\nu\pr)$ vs.~$\varepsilon_\nu\pr$ 
for $\nu_e n$ scattering for various $\varepsilon_\nu$]
{The $l=0$ term in the Legendre expansion of the neutrino-nucleon scattering
kernel, $\Phi_0(\varepsilon_\nu,\varepsilon_\nu\pr)$ (eq. (\ref{momentkernel})), for 
$T=6$ MeV and $\eta_n=-2$ as a 
function of $\varepsilon_\nu\pr$ for $\varepsilon_\nu=5$, 10, 15, 20, 25, and 35 MeV.
Note that for $\varepsilon_\nu=5$ MeV the neutrino is predominantly upscattered, while
for $\varepsilon_\nu=35$ MeV the neutrino is predominantly downscattered.  The magnitude 
of $\Phi_0(\varepsilon_\nu,\varepsilon_\nu\pr)$ and sign of $\langle\omega\rangle$
are to be compared with those in Fig. \ref{ek3}.}
\label{nk3}
\end{center}
\end{figure}

An important simplification comes from detailed balance, a consequence
of the fact that these scattering rates must drive the distribution to equilibrium.
One obtains: $R^{\rm in}=e^{-\beta\omega}R^{\rm out}$, where $\beta=1/T$.  Therefore, we need deal
only with $R^{\rm out}$.  The scattering kernels for 
inelastic neutrino-nucleon and neutrino-electron scattering
depend in a complicated fashion on scattering angle.  
For this reason, one generally approximates the angular dependence
of the scattering kernel with a truncated Legendre series (Bruenn 1985).  We take
\begin{equation}
R^{\rm out}(\varepsilon_\nu,\varepsilon_\nu\pr,\cos\theta)
=\sum_{l=0}^\infty\frac{2l+1}{2}\Phi(\varepsilon_\nu,\varepsilon_\nu\pr,\cos\theta)
P_l(\cos\theta),
\end{equation}
where
\begin{equation}
\Phi_l(\varepsilon_\nu,\varepsilon_\nu^\prime)=\int_{-1}^{+1}d(\cos\theta)\,
R^{\rm out}(\varepsilon_\nu,\varepsilon_\nu^\prime,\cos\theta)P_l(\cos\theta).
\label{momentkernel}
\end{equation}
Figure \ref{ek1} shows $R^{\rm out}(\varepsilon_\nu,\varepsilon_\nu\pr,\cos\theta)$
for $\nu$-electron scattering as a function of $\cos\theta$ for various values of
$\varepsilon_\nu^\prime$ at $\varepsilon_\nu = 20$ MeV and a representative thermodynamic point.
In practice, one expands only to first order so that
\begin{equation}
R^{\rm out}(\varepsilon_\nu,\varepsilon_\nu\pr,\cos\theta)\sim
\frac{1}{2}\Phi_0(\varepsilon_\nu,\varepsilon_\nu\pr)+
\frac{3}{2}\Phi_1(\varepsilon_\nu,\varepsilon_\nu\pr)\cos\theta.
\label{kernelapprox}
\end{equation}
Figure \ref{ek2} depicts the various Legendre contributions for the same point.
Figure \ref{ek3} provides the corresponding final-state energy distribution
function for various values of $\varepsilon_\nu$, using only the $l=0$ term.
Figures \ref{nk1} and \ref{nk3} are similar to Figs. \ref{ek1} and \ref{ek3},
but are for inelastic $\nu$-nucleon scattering at a representative thermodynamic point.

A simple approach to handling energy redistribution involves
calculating angle and energy moments and deriving source and sink terms 
(eqs. \ref{gencoll} and \ref{etatilde}; Thompson, Burrows, \& Pinto 2003).
Substituting into the first term on the right-hand-side of eq. (\ref{gencoll})
(the source) gives
\begin{equation}
\tilde{\eta}_\nu^{\rm scatt}=(1-f_\nu)
\int_0^\infty \frac{d\varepsilon_\nu\pr\varepsilon_\nu^{\prime\,2}}{c(2\pi\hbar c)^3}\,e^{-\beta\omega}
\int_{-1}^{+1}d\mu\pr f_\nu\pr\int_0^{2\pi}d\phi\pr
\left[\frac{1}{2}\Phi_0+
\frac{3}{2}\Phi_1\cos\theta\right]
\end{equation}
Substituting for $\cos\theta$ using eq. (\ref{costheta}) and using the definitions
\begin{equation}
\tilde{J}_\nu=\frac{1}{2}\int_{-1}^{+1}d\mu f_\nu
\label{jtilde}
\end{equation}
and
\begin{equation}
\tilde{H}_\nu=\frac{1}{2}\int_{-1}^{+1}d\mu \mu f_\nu
\label{htilde}
\end{equation}
we have that
\begin{equation}
\tilde{\eta}_\nu^{\rm scatt}=(1-f_\nu)\frac{4\pi}{c(2\pi\hbar c)^3}
\int_0^\infty d\varepsilon_\nu\pr \varepsilon_\nu^{\prime\,2} e^{-\beta\omega}
\left[\frac{1}{2}\Phi_0\tilde{J}_\nu\pr+\frac{3}{2}\Phi_1\mu\tilde{H}_\nu\pr\right].
\end{equation}
Integrating over $\mu$ to get the source for the zeroth moment of the transport equation,
\begin{equation}
\frac{1}{2}\int_{-1}^{+1}d\mu\,\tilde{\eta}_\nu^{\rm scatt}=
\frac{4\pi}{c(2\pi\hbar c)^3}
\int_0^\infty d\varepsilon_\nu\pr \varepsilon_\nu^{\prime\,2} e^{-\beta\omega}
\left[\frac{1}{2}\Phi_0\tilde{J}_\nu\pr(1-\tilde{J}_\nu)-\frac{3}{2}\Phi_1\tilde{H}_\nu\tilde{H}_\nu\pr\right].
\label{jtildeeta}
\end{equation}
Similarly, we can write the sink term of the Boltzmann equation  collision term 
(second term in eq. \ref{gencoll}), employing the Legendre expansion
\begin{equation}
\tilde{\chi}_\nu^{\rm scatt}=\frac{4\pi}{c(2\pi\hbar c)^3}
\int_0^\infty d\varepsilon_\nu\pr \varepsilon_\nu^{\prime\,2}
\left[\frac{1}{2}\Phi_0(1-\tilde{J}_\nu\pr)-\frac{3}{2}\Phi_1\mu\tilde{H}_\nu\pr\right].
\end{equation}
The contribution to the zeroth moment equation is then
\begin{equation}
\frac{1}{2}\int_{-1}^{+1}d\mu(-\tilde{\chi}_\nu^{\rm scatt}f_\nu)=-\frac{4\pi}{c(2\pi\hbar c)^3}
\int_0^\infty d\varepsilon_\nu\pr \varepsilon_\nu^{\prime\,2}
\left[\frac{1}{2}\Phi_0(1-\tilde{J}_\nu\pr)\tilde{J}_\nu-\frac{3}{2}\Phi_1\tilde{H}_\nu\tilde{H}_\nu\pr\right].
\label{jtildechi}
\end{equation}
Combining these equations, we find that
$$\frac{1}{2}\int_{-1}^{+1}d\mu\,{\cal L}^{\rm scatt}_\nu[f_\nu]=
\frac{4\pi}{c(2\pi\hbar c)^3}\int_0^\infty d\varepsilon_\nu\pr \varepsilon_\nu^{\prime\,2} \hspace{6cm}$$
\begin{equation}
\hspace*{1cm}\,\,\,\times\,\,\,
\left\{\frac{1}{2}\Phi_0\left[\tilde{J}_\nu\pr(1-\tilde{J}_\nu)e^{-\beta\omega}-(1-\tilde{J}_\nu\pr)\tilde{J}_\nu \right]
-\frac{3}{2}\Phi_1\tilde{H}_\nu\tilde{H}_\nu\pr(e^{-\beta\omega}-1)\right\}.
\end{equation}
One can see immediately that including another term in the Legendre expansion
(taking $R^{\rm out}\sim(1/2)\Phi_0+(3/2)\Phi_1\cos\theta+
(5/2)\Phi_2(1/2)(3\cos^2\theta-1)$) necessitates including $\tilde{P}_\nu$ and
$\tilde{P}_\nu\pr$, the second angular moment of the neutrino phase-space distribution
function, in the source and sink terms.  
While easily doable, we advocate retaining only
the linear term.

\section{Inelastic Neutrino Interactions with Relativistic Nucleons and Leptons}
\label{freegas}

We now explore more sophisticated formalisms for handling inelastic
scattering processes in nuclear matter.  In this section, we address the non-interacting nucleon case.

For neutrino energies of interest to supernova, which are less than a few
hundred MeV, we may write the neutrino-matter interaction  in terms of Fermi's effective Lagrangian
\begin{eqnarray}
{\cal L}_{int}^{cc} &=& \frac{G_F}{\sqrt{2}} ~~l_\mu
j_W^\mu\, \qquad {\rm ~for} \qquad \nu_l + B_2 \rightarrow  l + B_4 \\
{\cal L}_{int}^{nc} &=& \frac{G_F }{\sqrt{2}} ~~l_\mu^{\nu}
j_Z^\mu\, \qquad {\rm ~for} \qquad \nu_l + B_2 \rightarrow  \nu_l + B_4 \,,
\label{gsw}
\end{eqnarray}
where $G_F\simeq 1.436\times 10^{-49}~{\rm erg~cm}^{-3}$ is the Fermi weak coupling constant.  When the typical energy and momentum involved in the reaction are small (compared to the mass  and 1/size of the target particle)  the lepton and baryon weak charged currents are:
\begin{eqnarray}
l_\mu = {\overline \psi}_l \gamma_\mu
\left( 1 - \gamma_5 \right) \psi_\nu \,, \quad
j_W^\mu = {\overline \psi}_4 \gamma^\mu
\left( g_{V} - g_{A} \gamma_5 \right) \psi_2 \,.
\label{ccurrents}
\end{eqnarray}
Similarly, the baryon and neutrino neutral currents are given by
\begin{eqnarray}
l_\mu^{\nu} = {\overline \psi}_{\nu} \gamma_\mu
\left( 1 - \gamma_5 \right) \psi_\nu \,, \quad
j_Z^\mu =  {\overline \psi}_4 \gamma^\mu
\left( c_{V} - c_{A} \gamma_5 \right) \psi_2 \,,
\label{ncurrents}
\end{eqnarray}
where $2$ and $4$ are the baryon (or electron) initial state and final state labels, respectively (these are identical for neutral-current reactions). The vector and axial-vector coupling constants  ($c_V,g_V ~\&~c_{A},g_{A}$)  are  listed in Table~1 for the various charged- and neutral-current reactions of interest. The charged-current reactions are kinematically suppressed for  $\unu$ and $\tnu$ neutrinos. This is because their energy 
$E_{\nu_{\mu} / \nu_{\tau}} \simeq T \le m_{\mu}, m_{\tau}$.  On the other hand, neutral-current reactions are common to all neutrino species and the neutrino-baryon couplings are independent of neutrino flavor. Neutrino coupling to the lepton in the same family is modified since the scattering may proceed due to both $W$ and $Z$ exchange; the couplings shown in Table~1 reflect this fact.

\begin{table}[t]
\begin{center}
\begin{tabular}{|l|c|c|c|l|}
\hline
{Reaction } & $c_V,g_V$  & $c_A,g_A $ & $F_2$\\
\hline
$\nu_i + n \rightarrow \nu_i + n$ & $-1/2$ & $-g_A/2$ &
$- 2 \mu_n \sin^2\theta_W-\delta $\\
$\nu_i + p \rightarrow \nu_i+ p$ & $ 1/2-2\sin^2\theta_W$ & $g_A/2$ & $ - 2 \mu_p \sin^2\theta_W+\delta$\\
$\nu_{e} + e^- \rightarrow \nu_{e}+ e^-$ & $ 1/2+2\sin^2\theta_W$ & $-1/2$ & $0$\\
$\nu_{\mu/\tau} + e^- \rightarrow \nu_{\mu/\tau}+ e^-$ & $ -1/2+2\sin^2\theta_W$ & $-1/2$ & $0$\\
$\nu_e + n \leftrightarrow e^- + n $ & $1$ & $g_A$ & $2\delta$\\
\hline
\end{tabular}
\caption[coupling constants]
{Neutral and charged current vector, axial and tensor coupling constants. The axial coupling $g_A=-1.23$, the weak mixing angle $\sin^2\theta_W$=0.23. $\mu_n=-1.913$ and $\mu_p=1.793$ are the anomalous magnetic moments of the neutron and proton, respectively and $\delta=(\mu_p-\mu_n)/2=1.853$. }
\label{cvca}
\end{center}
\end{table}

From the structure of the current-current Lagrangian, we can calculate the differential cross section for neutrino scattering and absorption. We are generally interested in calculating scattering/absorption rates in matter. Hence, it is convenient to express results in terms of the differential scattering/absorption rate. For a neutrino with energy $E_1$, this is given by $d\Gamma(E_1)=\sum_{i}~c~d\sigma(E_1)_i/V$, where $d\sigma$ is the differential cross section, the sum is over the target particle in volume $V$, and $c=1$ is the (relative) velocity of the neutrinos. For reasons that will soon become clear,  it is useful to express the differential rate in terms of the current-current correlation function for the target particles (also called medium polarization tensor and is related to the dynamic structure factor) rather than the S-matrix element involving free asymptotic states.

The differential scattering rate of matter is given by  
(Horowitz \& Wehrberger 1991; Reddy et al. 1998): 
\begin{eqnarray}
\frac {d\Gamma(E_1)}{d^2\Omega_3 dE_3}=  -\frac {G_F^2}{32\pi^2}~\frac{E_3}{E_1}
~\frac{[1-f_3(E_3)]}{\left[1-\exp{\left(\frac{-\omega-(\mu_2-\mu_4)}{T}\right)}\right]} {\rm Im}~\left[L^{\alpha\beta}\Pi^R_{\alpha\beta}\right] \,,
\label{dcross1}
\end{eqnarray}
where the incoming neutrino energy is $E_{1}$, the outgoing neutrino (or electron) energy is $E_{3}$, and $\omega = E_1 - E_3$. The factor $[1-\exp((-\omega-\mu_2+\mu_4)/T)]$ arises due to the fluctuation-dissipation theorem, since particles labeled `2' and `4' are in thermal equilibrium at temperature $T$ and in chemical equilibrium with chemical potentials $\mu_2$ and $\mu_4$, respectively. For neutral-current processes, $\mu_4=\mu_2$, and for charged-current processes, such as $\nu_e + n \rightarrow e^- + p$, $\mu_2=\mu_n$ and $\mu_4=\mu_p$. The final-state Pauli blocking of the outgoing lepton is accounted for by the factor $(1-f_3(E_3))$. The lepton tensor $L_{\alpha\beta}$ is given by
\begin{equation}
L^{\alpha\beta}= 8[2k^{\alpha}k^{\beta}+(k\cdot q)g^{\alpha\beta}
-(k^{\alpha}q^{\beta}+q^{\alpha}k^{\beta})\mp i\epsilon^{\alpha\beta\mu\nu}
k^{\mu}q^{\nu}]\,,
\end{equation}
where the sign of the last term is positive for anti-neutrinos and negative for neutrinos. 

The medium or target particle retarded polarization tensor is given by
\begin{equation}
{\rm Im} \Pi^R_{\alpha\beta} =
\tanh{\left(\frac{\omega+(\mu_2-\mu_4)}{2T}\right)} {\rm Im}~\Pi_{\alpha\beta}
\,,\\
\end{equation}
where $\Pi_{\alpha\beta}$ is the time-ordered or causal polarization and is given by
\begin{equation}
\Pi_{\alpha\beta}=-i \int
\frac{d^4p}{(2\pi)^4} {\rm Tr}~[T(G_2(p)~\Gamma_{\alpha} ~G_4(p+q)~\Gamma_{\beta})]\,.
\label{polarization}
\end{equation}
Above, $k_{\mu}$ is the incoming neutrino four-momentum and $q_{\mu}$ is  the four-momentum  transfer. In writing the lepton tensor, we have neglected the electron mass term, since typical electron energies are of the order of a 
hundred MeV at high densities.  The Green's functions $G_i(p)$ (the index $i$
labels particle species) describe the propagation of baryons at finite
density and temperature.  In a Fermi gas, the fermion Green's function is given by
\beq
G(p) = \frac{1}{\not \!p+\mu \gamma_0-m}\,,
\eeq 
where $\mu$ is the chemical potential and $m$ is the mass. For the neutral (charged) current, the vertex function $\Gamma_{\mu}$ is $c_V (g_V) \gamma_{\mu}$
for the vector current and $c_A (g_A) \gamma_{\mu}\gamma_5$ for the axial current. Given
the structure of the particle currents, we have
\begin{eqnarray}
\Pi_{\alpha\beta}
&=&c_V^2~\Pi_{\alpha\beta}^{V}~+~c_A^2~\Pi_{\alpha\beta}^{A}
~+~2c_Vc_A~\Pi_{\alpha\beta}^{VA} \,.
\label{polarform}
\end{eqnarray}
For the neutral-current, the vector polarization $\Pi^V_{\alpha,\beta}$ is obtained by substituting $\{\Gamma_\alpha,\Gamma_\beta\}::\{\gamma_\alpha,\gamma_\beta\}$;
the axial polarization $\Pi^A_{\alpha,\beta}$ is obtained by substituting  $\{\Gamma_\alpha,\Gamma_\beta\} ::
\{\gamma_\alpha\gamma_5,\gamma_\beta\gamma_5\}$;  and the mixed
part $\Pi^{VA}_{\alpha,\beta} $ is obtained by substituting $\{\Gamma_\alpha,\Gamma_\beta\} ::\{\gamma_\alpha\gamma_5,
\gamma_\beta\}$. 

Using vector current conservation and translational
invariance, $\Pi_{\alpha\beta}^V$ may be written in terms of two independent
components. In a frame where $q_{\mu}=(\omega,|q|,0,0)$, we have
\begin{eqnarray*}
\Pi_T = \Pi^V_{22} \qquad {\rm and} \qquad
\Pi_L = -\frac{q_{\mu}^2}{|q|^2}\Pi^V_{00} \,.
\end{eqnarray*}
The axial current-current correlation function can be written as a vector
piece plus a correction term:
\begin{eqnarray}
\Pi_{\mu \nu}^A=\Pi^V_{\mu \nu}+g_{\mu \nu}\Pi^A \,.
\end{eqnarray}
The mixed, axial-current/vector-current correlation function is
\begin{eqnarray}
\Pi_{\mu \nu}^{VA}= i\epsilon_{\mu, \nu,\alpha,0}q^{\alpha}\Pi^{VA}\,.
\end{eqnarray}
These simplifications enable us to write the differential rate in eq. (\ref{dcross1}), as follows
\begin{eqnarray}
\frac{d\Gamma}{d^2\Omega~dE_{3}} &=& \frac{G_F^2}{4
\pi^3}~ \frac {E_3}{E_1} q_{\mu}^2~\left[1-f_3(E_3)\right]~ \nonumber \\
&\times& [A~S_1(|\vec{q}|,\omega)~+
~S_2(|\vec{q}|,\omega)~\pm B~S_3(|\vec{q}|,\omega)  ] \,,
\nonumber \\
A&=&\frac{4E_1E_3+q_{\mu}^2}{2q^2} \,, \quad B= E_1+E_3 \,,
\label{dcross2}
\end{eqnarray}
where the plus is for neutrinos and the minus is for anti-neutrinos. The three response functions, $S_1$, $S_2$, and $S_3$ are given by
\begin{eqnarray}
S_1(|\vec{q}|,\omega)&=&\frac{({c_V}^2+{c_A}^2) ~\left(
{\rm Im}~\Pi^R_L(q,\omega)+{\rm Im}~ \Pi^R_T(q,\omega)\right)}
{1-\exp{\left(-\beta(\omega+\mu_2-\mu_4)\right)}}\\
S_2(|\vec{q}|,\omega)&=&\frac{({c_V}^2 + {c_A}^2)~{\rm Im}~ \Pi^R_T(q,\omega)- c_A^2~{\rm Im}~ \Pi^R_A(q,\omega)}{1-\exp{\left(-\beta(\omega+\mu_2-\mu_4)\right)}}\\
S_3(|\vec{q}|,\omega)&=&\frac{2{c_V}{c_A} ~{\rm Im}~ \Pi^R_{VA}(q,\omega)}{1-\exp{\left(-\beta(\omega+\mu_2-\mu_4)\right)}}  \,.
\label{resp}
\end{eqnarray}
The imaginary parts of the polarization functions needed to evaluate the response functions have been calculated explicitly by Horowitz \& Wehrberger (1991,1992) and Reddy et al. (1999). Neutrino scattering and absorption kinematics probes space-like kinematics where $q_{\mu}^2\le0$ (this is true only for massless neutrinos and electrons; however, when the typical lepton energies are large compared to $m_e$, restriction to space-like kinematics is a good approximation). For space-like excitations, 
\begin{eqnarray}
{\rm Im}~ \Pi^R_L(q,\omega)&=&\frac{q_{\mu}^2}{2\pi |q|^3}
\left[I_2 + \omega I_1+\frac{q_{\mu}^2}{4} I_0\right] \\
{\rm Im}~ \Pi^R_T(q,\omega)&=&\frac{q_{\mu}^2}{4\pi |q|^3}
\left[I_2 + \omega I_1+
\left(\frac{q_{\mu}^2}{4} +
\frac{q^2}{2}+M_2^{^2}\frac{q^2}{q_{\mu}^2}\right) I_0\right] \\
{\rm Im}~ \Pi^R_A(q,\omega)&=&\frac{M_2^{^2}}{2\pi |q|}I_0\\
{\rm Im}~ \Pi^R_{VA}(q,\omega)&=&\frac{q_{\mu}^2}{8\pi |q|^3}
[\omega I_0+2I_1] \,,
\label{polar}
\end{eqnarray}
where $M_i$ is the rest mass of particle $i$.
The one-dimensional integrals
\begin{equation}
I_n=\tanh{\left(\frac{\omega+(\mu_2-\mu_4)}{2T}\right )}\int_{e_-}^{\infty}dE~E^n~[F(E,E+\omega)+F(E+\omega,E)]\,, \\
\end{equation}
where $F(x,y)=f_2(x)(1-f_4(y))$ and $f_i(E)$ are Fermi-Dirac particle distribution functions.  The lower limit on the integral, $e_-$, arises due to kinematical restrictions and is given by
\begin{eqnarray}
e_-&=&-\beta \frac{\omega}{2}
+\frac{q}{2}\sqrt{\beta^2-4\frac{M_2^{2}}{q^2-\omega^2}}\,,
\end{eqnarray}
where
\begin{equation}
\beta = 1+\frac{M^{^2}_4-M^{^2}_2} {q^2-\omega^2}\,.
\end{equation}
These integrals may be explicitly expressed in terms of the Polylogarithmic
functions
\begin{equation}
Li_n(z) = \int_0^z \frac {Li_{n-1}(x)}{x} \,dx \,,
\qquad Li_1(x) = \ln (1-x) \, .
\end{equation}
This Polylogarithm representation is particularly useful and compact:
\begin{eqnarray}
I_0 &=& T~z\left(1+\frac{\xi_1}{z}\right)\,, \\
\label{iis0}
I_1 &=& T^2~z\left(\frac{\mu_2-U_2}{T}  -
\frac{z}{2}+\frac{\xi_2}{z}+\frac{e_-\xi_1}{zT}\right) \,,\\
\label{iis1}
I_2 &=& T^3~z~\left(\frac{(\mu_2-U_2)^2}{T^2} -
z\frac{\mu_2-U_2}{T} +
\frac{\pi^2}{3}+\frac{z^2}{3}-2\frac{\xi_3}{z} + 2\frac{e_-\xi_2}{Tz}
+\frac{e_-^2\xi_1}{T^2z}\right) \,, \nonumber \\
\label{iis2}
\end{eqnarray}
where $z=(\omega+(\mu_2-\mu_4))/T$ and the factors $\xi_n$ are given by
\begin{equation}
\xi_n = Li_n(-\alpha_1) - Li_n(-\alpha_2)\,,
\label{dlis}
\end{equation}
with
\begin{equation}
\alpha_1 = \exp\left((e_--\mu_2)/T\right)\,,\quad
\alpha_2 = \exp\left((e_-+\omega-\mu_4)/T\right) \,.
\end{equation}
In the case of neutral currents, some of the terms above simplify
\begin{eqnarray}
z = \frac{\omega}{T} \,, \quad \mu_2 = \mu_4\,, \quad
e_- = -\frac{\omega}{2}
+\frac{q}{2}\sqrt{1-4\frac{M_2^{*2}} {q_{\mu}^2}}\, .
\end{eqnarray}

The total scattering rate is the double integral in $(q,\omega)$ space:
\begin{eqnarray}
\Gamma=
\frac{G_F^2}{2\pi^2E_1^2} \int_{-\infty}^{E_1}
d\omega~ (1-f_3(E_3))
\int_{|\omega|}^{2E_1-\omega} dq~q~q_{\mu}^2~ [AS_1+S_2 \pm B S_3]\,.
\label{master}
\end{eqnarray}
Eq.(\ref{master}) allows us to calculate the cross section per unit volume, or equivalently the inverse mean free path. This naturally
incorporates the effects of Pauli blocking, Fermi and thermal motion
of target particles, and the contribution of relativistic terms to the electron and baryon currents. The formalism reviewed in this section provides a unified description of reactions (both charged and neutral currents) involving both nucleons and electrons. In the absence of strong and electromagnetic interactions, this is almost all there is to the subject of neutrino opacities in matter with nucleons and leptons. The only omission here is the effect of weak magnetism (\S\ref{CCabs}). This arises due to tensor coupling between neutrinos and nucleons and is discussed next.

The vector and axial-vector structure of the weak hadronic currents given in eqs. (\ref{ccurrents}) and (\ref{ncurrents}) are valid only at low energy ($E_\nu \ll M$). When the neutrino energy becomes comparable to the intrinsic energy scale that characterizes the hadron (its size and mass) additional contributions arise. These include form factors and the tensor (weak magnetism) couplings. To leading order in the four-momentum transfer expansion ($q_{\mu}$), the hadronic neutral and charged weak currents take the form
\begin{eqnarray}
j_W^{\mu} &=& {\overline \psi}_4
\left[ g_{V}(q^2_{\mu})  \gamma^\mu+ i F_2(q^2_{\mu}) \sigma^{\mu\nu}\frac{q_{\nu}}{2M}
- g_{A}(q^2_{\mu}) \gamma^\mu \gamma_5 \right] \psi_2 \, \\
j_Z^\mu &=& {\overline \psi}_4 
\left[c^{i}_{V}(q^2_{\mu})\gamma^\mu -i F^i_2(q^2_{\mu}) \sigma^{\mu\nu}\frac{q_{\nu}}{2M}- c^{i}_{A} (q^2_{\mu}) \gamma^\mu
\gamma_5 \right] \psi_2 \,.
\end{eqnarray}
The $q^2_{\mu}$ dependence of the vector and axial couplings is weak and one can safely ignore these corrections, but the contribution from weak magnetism is not negligible (Vogel 1984). The tensor coupling constants for the various reactions of interest are given in Table 1.   For a neutrino of energy $E_\nu \simeq 50$ MeV, a typical $|\vec{q}| \sim 2E_\nu \sim 100$ MeV, the weak magnetism correction to the vector current is then 
$F_2 |\vec{q}|/(2M) ~\simeq 6 \% $.   Perhaps, more importantly, these corrections appear with opposite sign for neutrinos and anti-neutrinos (Vogel 1984; Horowitz 2002), resulting in $10-20\%$  
differences in the relative cross sections. 

Tensor coupling introduces new structure to the current-current correlation function given in eq. (\ref{polarform}) and corresponding changes to the response functions given in eq. (\ref{resp}). These corrections have been computed in recent work by \cite{horowitz03}. To order $q^2/M$, the tensor-vector, tensor-axial vector and tensor-tensor polarization functions contribute to the neutrino response. These are obtained from eq. (\ref{polarization}), with the following substitutions: for the tensor-vector, $\Gamma_\alpha=iF_2\sigma_{\alpha,\delta}q^\delta/2M$ and $\Gamma_\beta=c_v\gamma_\beta$; for the tensor-axial, $\Gamma_\alpha=iF_2\sigma_{\alpha,\delta}q^\delta/2M$ and $\Gamma_\beta=c_a\gamma_\beta\gamma_5$; and for the tensor-tensor, 
$\Gamma_\alpha=iF_2\sigma_{\alpha,\delta}q^\delta/2M$ and $\Gamma_\beta=iF_2\sigma_{\beta,\delta}q^\delta/2M$. These polarization functions can be related to the vector, axial, and vector-axial polarization tensors
discussed earlier. Explicit forms for the imaginary parts of these polarization tensors can be found in \cite{horowitz03}.
\section{Strong and electromagnetic correlations between nuclei, nucleons, and electrons} 
\label{strongandeandm}

The results of the previous sections provide a complete description of neutrino cross sections in an ideal Fermi gas of nucleons and leptons. Here, we address non-ideal corrections to neutrino scattering rates arising due to interactions between nucleons, electrons and nuclei. The role of correlations in neutrino opacities was first appreciated in the pioneering works of \cite{sawyer75} and \cite{iwamoto82}. They showed that strong interactions between nucleons could significantly modify (by as much as a factor of 2-3) neutrino cross sections at densities and temperatures of relevance to supernovae. In the past decade, there have been several attempts to incorporate these effects in calculations of the neutrino opacity of dense matter. While much progress has been made in understanding qualitative aspects of these corrections, quantitative predictions for the neutrino cross sections at high density remain elusive. We present a pedagogic review of the qualitative findings of model calculations, provide an overview of the current state of the art of many-body effects in neutrino cross sections, and comment on their shortcomings.  

\subsection{Plasma of Heavy Ions: Neutrino Opacity at $\rho \sim 10^{12}$ g/cm$^3$: }
\label{ionplasma}
The simplest system, with non-trivial many-body dynamics, which is relevant in the supernova context is a plasma of heavy nuclei (like Fe) immersed in a background of degenerate electrons. \cite{freed} showed that the dominant source of opacity for neutrinos in such a plasma is the coherent scattering off nuclei.  Low-energy neutrinos can couple coherently, via the vector neutral current, to the total weak charge of the nucleus. The weak charge of a nucleus with $A$ nucleons and $Z$ protons is given by $Q_{W}= ( 2Z- A - 4 Z\sin^{2}\theta_{W})/2$. 

For neutrino-nucleus scattering in a plasma, we must properly account for the presence of other nuclei, since scattering from these different sources can interfere. In the language of many-body theory, this screening is encoded in the density-density correlation function. It is therefore natural to express the 
scattering rate in terms of these correlation functions as described earlier in  \S\ref{freegas}.   To motivate the relation between the cross section and the  density-density correlation function we begin by noting that the effective Lagrangian describing the neutral-current interaction of low-energy neutrinos with nuclei is given by 
\beq
L_{\mathrm{NC}}=\frac{G_{\mathrm{F}}}{\sqrt{2}}~Q_W ~l_{\mu}~j^{\mu}
\label{leff1}
\eeq
where $l_{\mu}=\overline{\nu}\gamma_{\mu}(1-\gamma_{5})\nu$ is the neutrino neutral current. Nuclei are heavy, and, correspondingly, their thermal velocities are small ($v \cong \sqrt{T/M} \ll c$). For simplicity, we assume that nuclei are bosons characterized only by their charge and baryon number.
In this case it is an excellent approximation to write the neutral current carried by the nuclei as  $j^{\mu}=\psi^{\dag}\psi~\delta^{\mu}_{0}$. We can write the differential scattering rate in terms of the density operator in momentum space
given by 
\beq 
\rho(\vec{q},t)= \psi^{\dag}\psi=\sum_{i=1\cdots N}~\exp(i\vec{q}\cdot\vec{r}_{i}(t))\,, 
\eeq
where the sum is over $N$ particles in a volume $V$ which are labeled  $i=1\cdots N$. The rate for scattering of a neutrino with energy $E_\nu$ to a state with energy $E'_{\nu}=E_\nu-\omega$, at an angle $\theta$, with a momentum transfer $\vec{q}$ is given by 
\begin{equation}
\frac{d\Gamma}{d\cos\theta dE'_\nu}=\frac{G^2_{\mathrm{F}}}{4\pi^2}~Q^2_W~(1+\cos\theta)~{E'_\nu}^2 ~S(|\vec{q}|,\omega) \,.
\end{equation}
The function $S(|\vec{q}|,\omega)$ is called the dynamic structure function. It embodies all spatial and temporal correlations between target particles arising from strong or electromagnetic interactions and is given by
\beq
S(|\vec{q}|,\omega)=  \int^{\infty}_{-\infty} dt ~ \exp(i\omega t) 
~\langle \rho(\vec{q},t)\rho(-\vec{q},0)\rangle \,.
\label{sofqw}
\eeq
In the above, $\langle \cdots \rangle$ denotes the ensemble average per unit volume. Note that $S(|\vec{q}|,\omega)$ is normalized such that in the elastic limit for a non-degenerate, non-interacting gas  $S(|\vec{q}|,\omega) = 2\pi~\delta(\omega)~n$, where $n=N/V$ is the density of particles.

To calculate the structure function we need to solve for the dynamics ($\vec{r}(t)$) of the ions as they move in each others presence, interacting via the two-body ion-ion interaction potential. The electrons are relativistic and degenerate,  and the electron Fermi momentum $k_{\mathrm{F}}=(3\pi^2~Z~n)^{1/3}~$. The timescale associated with changes in their density distribution is rapid compared to the slow changes we expect in the density field of the heavy ions.  Electrons almost ``instantaneously" follow the ions and screen the Coulomb potential between ions. Correspondingly, there is small excess in the electron distribution around each ion.  As first discussed by Leinson et al. (1988), this excess will also screen the weak charge of the nucleus to which the neutrinos couple. Electronic screening was discussed earlier in \S\ref{nuA} and eq. (\ref{nunucleuscs}).  Here, we focus on the dynamics of ions interacting through the screened potential $V(r) = Z^2 e^2 \exp(-r/\lambda_e)/(4 \pi r)$, where $\lambda_e=\sqrt{4 \alpha_{\mathrm{em}} / \pi}~
k^{-1}_{\mathrm{F}}$ is the electron Debye screening length.  

At $10^{12}$ g/cm$^3$, the typical inter-ion distance is $d \lesssim 30$ fm, $\lambda_e \ge 80$ fm, and the typical ion-ion interaction energy, $E_{\mathrm{pot}}\simeq Z^2 e^2/(4 \pi d)$, is large. For temperature in the range $1-5$ MeV, the ratio of the potential energy to the kinetic energy, $\Gamma$, equals $Z^2 \alpha_{\mathrm{em}}/(d ~kT)$. For nickel-like nuclei with $Z=28$, $\Gamma=8.8$ for $T=4$ MeV. The dynamics of such a strongly coupled plasma is not amenable to analytic methods of perturbation theory and approximate non-perturbative methods have had limited success in describing these systems.  For a classical system of point particles interacting via a 2-body potential it is possible to numerically simulate the real-time dynamics of the system.  Such simulations confine  N particles to a box with periodic boundary conditions,  calculate the force on each particle at any time, and evolve the particles by using their equations of motion ($\ddot{\vec{r}}=\mathrm{Force}/\mathrm{M}$, where $M$ is the mass of the ion). This technique, which goes by the name molecular dynamics (MD), has been used extensively in condensed matter physics, plasma physics, and chemistry.  For an early application of this technique to the study of the response of a one-component plasma see \cite{Hansen:jp}. 

\begin{figure}
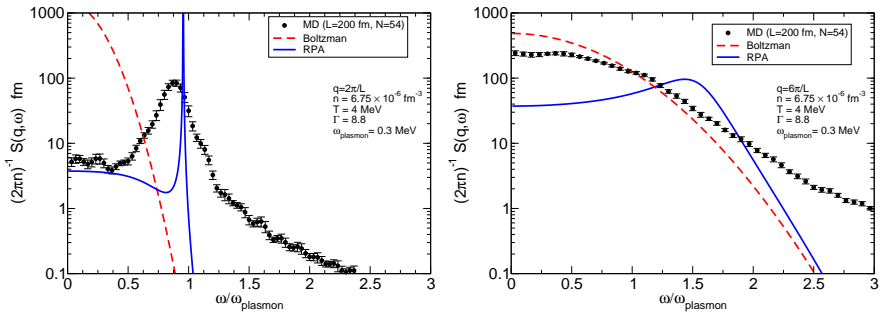

\begin{center}
\resizebox{.48\textwidth}{!}{%
\includegraphics{bstfig6.1.eps}
}
\resizebox{.48\textwidth}{!}{%
\includegraphics{bstfig6.2.eps}
}
\caption{Dynamic structure function of a plasma of ions as a function of energy transfer $\omega$ (measured in units of the plasma frequency $\omega_p\simeq0.3$ MeV) and fixed momentum transfer
$|\vec{q}|=2\pi/L\simeq6$ MeV (left panel) and
$|\vec{q}|=6\pi/L\simeq18$ MeV (right panel). Statistical errors 
for the results of the molecular dynamics simulations are also indicated. }
\label{MD}
\end{center}
\end{figure}

The de Broglie wavelength of the ions is $\Lambda_D \simeq 4.6\times A^{-1}~T^{-1}_{\mathrm{MeV}}$ fm, where $A$ is the mass number of the ion and $T_{\mathrm{MeV}}$ is the temperature in MeV. For $T\ge 1$ MeV, $\Lambda_D \ll d$, the  ion-plasma  is classical, and we may implement MD to calculate $S(|\vec{q}|,\omega)$.  Fig. (\ref{MD}) shows the results of such a calculation (from Luu et al. 2004).  We chose to simulate 54 ions (A=56,Z=28) in a box of length $L=200$ fm. This corresponds to an ion density $n=6.75\times 10^{-6}$ fm$^{-3}$ or the mass density $\rho\approx 6 \times 10^{12}$ g/cm$^3$. The background electron density was chosen to make the system electrically neutral. For a classical simulation, the system is characterized by the plasma $\Gamma$ defined earlier. Figure (\ref{MD}) shows a comparison between the results for $S(\vec{q},\omega)$  obtained by MD simulations (dots with error bars), the free (Boltzmann) gas response (dashed-line), and the response obtained using Random Phase Approximation (RPA). The response in RPA is discussed in the next section.  The results are shown as a function of  $\omega/\omega_\mathrm{plasmon}$, where $\omega$ is the energy transfer, and $\omega_\mathrm{plasmon}=\sqrt{4 \pi~\alpha_{\mathrm{em}}~n/M}$ is the plasma frequency of the classical heavy-ion plasma. 

\subsubsection{ Random Phase Approximation (RPA)}

We motivate and present a heuristic derivation of the structure function calculated using the random phase approximation. To begin, we note that the dynamic structure factor for a non-relativistic and non-interacting  Maxwell-Boltzmann gas with a chemical potential $\mu_A$ and number density $n$  is given by
\begin{eqnarray}
S_{\mathrm{MB}}(\vec{q},\omega)&=&~\int~\frac{d^3p}{(2\pi)^3}~f(E_{\vec{p}})~
2\pi~\delta\left(E_{\vec{p}}-E_{\vec{p}+\vec{q}} - \omega\right) \nonumber \\
&=&  ~\frac{n}{q}~\sqrt{\frac{2 \pi~M} {kT}}~\exp(-Q_+^2)\,, \\
\mathrm{where}~Q_+^2~&=& ~ \frac{M}{2kT}\left( -\frac{\omega}{q} +\frac{q}{2M} \right)^2 \,, \label{Qplus}\\
\mathrm{and}~n~&=&\left(\frac{M~kT}{2\pi}\right)^{3/2}~\exp\left(\frac{\mu_A}{kT}\right)\,.
\label{ndnrsofqw}
\end{eqnarray}
This follows directly from the equation of motion of the free particles and eq. (\ref{sofqw}).  When an external current couples to the interacting system at wavelengths large compared to the inter-particle spacing, scattering amplitudes from different particles can interfere. In the case of neutrino scattering this leads to screening of the weak charge. RPA is known to provide an adequate  description of this dynamic screening in the long-wavelength limit.  Pioneering work by \cite{bohm} established RPA as a useful non-perturbative approximation. It is used widely in nuclear and solid state physics to describe the long-wavelength response of many-particle systems. In particular, RPA accounts for dynamic screening and is able to correctly predict the presence of collective modes.  These correspond to a plasmon in the case of unscreened Coulomb interactions and phonons in the case of screened interactions (as in the present case). The RPA results in Fig. \ref{MD} clearly show the presence of these collective states.

Screening in RPA is accounted for by the selective re-summation of an infinite series of bubble graphs. The bubbles are free excitations and are connected by the two-body interaction potential. The lower diagram in Fig. \ref{schwingerdyson} represents this series. The solid lines represent propagation of ions, the wavy line represents the external current (the neutral current), and  the dashed lines represent the screened Coulomb interaction.  The diagram illustrates screening of the coupling between the external probe (neutrinos in the present case) and particles in the plasma through the intermediate particle-hole excitations.  The bubble-sum is a geometric series in the interaction and can be evaluated analytically for simple interactions. For the ion-plasma, the dynamic structure function in RPA is given by
\begin{eqnarray}
S_{\mathrm{RPA}}(\vec{q},\omega)&=&~
\frac{2\hbar}{1-\exp \left(-\beta~\hbar\omega\right)}~ 
\mathrm{Im}~
\left[ \frac{\Pi_0 (\vec{q},\omega) } {1-V_{\mathrm{C}}(q)~\Pi_0(\vec{q},\omega)} \right] \,, \\
\mathrm{where}~V_{\mathrm{C}}(q)&=&\alpha_{\mathrm{em}}~Z^2~\left(\frac{\lambda_e^2}{1+q^2\lambda_e^2}\right)\,.
\end{eqnarray}
The expression for $V_{\mathrm{C}}(q)$ is valid for $q \ll k_{\mathrm{F}}$, where  $\lambda_e$ is the Debye screening length and $k_{\mathrm{F}}$ is the Fermi momentum of the background degenerate electron gas. The free-gas polarization function
\beq
\Pi_0(\vec{q},\omega)=~i~\int \frac{d^4p}{(2\pi)^4}~G(p)~G(p+q)\,,
\eeq
where $G(p)=(p_0-(p^2/2M-\mu_A) )^{-1}$ is the free particle, non-relativistic, Green's function and $\mu_A$ is chemical potential of the ions. Analytic expressions for the real and imaginary parts of $\Pi_0$ were first obtained by \cite{lindhard}. In the case of a Boltzmann gas these are given by 
\begin{eqnarray}
\mathrm{Im}~\Pi_0(\vec{q},\omega) &=&~\frac{\beta~M^2}{2\pi ~q}~\left(1-\exp\left(-\beta~\hbar\omega\right)\right)~\exp(-Q_+^2) \,, \\
\mathrm{Re}~\Pi_0(\vec{q},\omega) &=&   \mathcal{P}~\int~\frac{d\omega'}{\pi}~\frac{\mathrm{Im}~\Pi_0(\vec{q},\omega')}{\omega-\omega' }\,,
\end{eqnarray}
where $Q_+^2$ was defined earlier in eq. (\ref{Qplus}).  

The comparison in Fig. \ref{MD} shows that both RPA and the free-gas response differ quantitatively. MD is exact in the classical limit. Corrections to classical evolution are governed by the smallness of the expansion parameter $\Lambda_D/d$. In the present case $(\Lambda_D/d) \ll 1\%$. In the left panel, the response for long wavelengths ($|\vec{q}|=2\pi/L\simeq6$ MeV) is shown. In this case,  the discrepancy between RPA and MD is not significant, supporting the expectation that RPA provides a fair description at long wavelengths. It captures, albeit much too sharply, the phonon peak (which in solid state parlance is called the Bohm-Staver sound) seen in MD simulations. The right panel shows results for $q=6\pi/L\simeq18$ MeV. This corresponds to a wavelength that is comparable to the inter-ion distance. Here, the plasmon mode is damped in RPA by single-particle excitations, since $\omega_{\mathrm{plasmon}} \simeq q v_\mathrm{thermal}$, where $v_\mathrm{thermal}=\sqrt{3~kT/M}$ is the thermal velocity of the ions. 
Nonetheless, this damping is still too weak to dissolve the plasmon peak. In contrast, results obtained using MD do not show any collective behavior, suggesting that RPA underestimates damping.
   
These results clearly illustrate two important features of the response of strongly-correlated systems: (1) correlations can greatly affect the shape of the low-energy response; and (2)  approximate many-body methods such as RPA can capture qualitative aspects of the response at large wavelength, but do poorly even at moderate wavelength. The failure of RPA can be attributed to the neglect of collisional damping from the excitation of multi-particle states. As the ions move, they ``simultaneously" scatter off several correlated neighboring particles. In RPA, these correlations are neglected. Ignoring these multi-pair excitations is a reasonable approximation in the long-wavelength limit. We will justify this in our subsequent discussion of sum rules (\S\ref{sumrules}), where we will argue that the spectrum of multi-pair excitations vanishes as $q^2$ in the $q\rightarrow 0$ limit.  However, the results indicate significant contribution from multi-pair excitations even at moderate wavelength. Finding ways to incorporate collisional damping of single-particle motion in the response has a long history in solid state physics, starting with the pioneering work of \cite{kadanoff}. We will not review these techniques here. Our focus was to illustrate the importance of many-particle dynamics and provide a baseline for comparison.  The lessons learned here will prove valuable in gauging the validity of results that will discussed in \S\ref{nuclearmatter}, where we evaluate the results obtained using RPA.
 
\subsubsection{Sum Rules}
\label{sumrules}

The dynamic structure factor is subject to constraints arising  due to thermodynamic considerations and symmetries of the Hamiltonian (conservation laws) describing the many-particle system. These constraints, called sum rules, provide valuable guidance in developing approximation schemes needed to compute  $S(|\vec{q}|,\omega)$. In the following, we discuss two such sum rules: (i) the compressibility sum rule; and (ii) the energy-weighted sum rule. 

The compressibility sum rule is related to the long-wavelength limit of the static structure function which is defined as 
\beq
S_{|\vec{q}|}=\frac{1}{n}~\int^{\infty}_{-\infty} ~\frac{d\omega}{2\pi}~S(|\vec{q}|,\omega)\,.
\eeq 
In the long-wavlength limit ($\vec{q} \rightarrow 0$), the static structure function is related to the isothermal compressibility by the following relation
\begin{eqnarray}
\lim_{\vec{q} \rightarrow 0} ~S_{|\vec{q}|}&=& \frac{1}{2\pi~n}~\lim_{\vec{q} \rightarrow 0} ~\langle \rho(-\vec{q},0) \rho(\vec{q},0) \rangle \nonumber \\
&= &n~kT~\mathcal{K}_T \,,\\
\mathrm{where}\quad \mathcal{K}_T&=& -\frac{1}{V}~\left(\frac{\partial V}{\partial P}\right)_{T}
\label{compressibility}
\end{eqnarray}
is the isothermal compressibility and $n$ is the particle number density. The first of the above relations follows directly from the  definition of $S(q,\omega)$ in terms of the density-density correlation function in eq. (\ref{sofqw}).  The above relation permits us to verify consistency between the neutrino response functions and the equation of state.     

The energy-weighted or F sum rule for any operator $\hat{\mathcal{O}}$ is given by
\beq 
\int^{\infty}_{0} ~\frac{d\omega}{2\pi}~\omega~
\left(\frac{1-\exp \left(-\beta~\hbar\omega\right)}{\hbar}\right)~
S_{\hat{\mathcal{O}}}(\vec{q},\omega) = \langle~\left[ [H,\hat{\mathcal{O}}(\vec{q})],\hat{\mathcal{O}}(\vec{q}) \right]~\rangle
\label{esum}
\eeq
In the absence of any velocity-dependent terms in the Hamiltonian, and for the specific case of the density operator, the right hand of  eq. (\ref{esum}) is independent of interactions and is given by
\beq
\langle~\left[[H, \rho(\vec{q})],\rho(\vec{q})\right]~\rangle=n~\frac{q^2}{2M} \,.
\eeq
Note that the F sum rule vanishes in the long-wavelength limit when the operator $\rho(0)$ commutes with the Hamiltonian. This is a consequence of the conservation of particle number and provides a useful constraint for the large $\omega$ behavior of $S(\vec{q},\omega)$ when it is computed using approximate many-body methods. For the  response of the ion-plasma discussed above, the results obtained in the case of the free Boltzmann response, the RPA, and molecular dynamics satisfy this F sum rule. Despite the large differences in the low-energy strength. Another important consequence of the F sum-rule is that it requires response at long wavelengths to vanish as $q^2$. This implies that multi-particle excitations are suppressed at long wavelength when the operator commutes with the Hamiltonian.  It also offers an explanation for the trend seen in Fig. \ref{MD}, where RPA results improve at large wavelength. 

\subsection{Pasta Phases: Neutrino Opacity at $\rho \ge 10^{13}$ g/cm$^3$}
\label{pasta}
With increasing density, the nuclei get bigger and the internuclear distance becomes smaller.  Under these conditions, the nuclear surface and Coulomb contributions to the free energy of the system become important. Ravenhall, Pethick, \& Wilson (1983) showed that it is energetically favorable for nuclei to deform  and assume novel, non-spherical, shapes such as rods and slabs. Further, the energy differences between these various shapes are small $\Delta E \simeq 10-100 $ keV. The dynamics of such an exotic heterogeneous phase is a complex problem involving several energy scales. Both strong and electromagnetic interactions play an  important role. For temperatures of interest, $T\lesssim 5$ MeV, the de Broglie wavelength and the inter-particle distance are comparable and quantum effects cannot be neglected.  Recently, 
\cite{Watanabe:2003xu} have studied the behavior of such matter using the techniques of  quantum molecular dynamics and also find rod- and slab-like configurations. How does the heterogeneity and existence of several low-energy excitations involving shape fluctuations influence the response of this phase to neutrinos?  In the simplest description,  the structure size ($r$) and the inter-structure  distance ($R$) characterize the system. We can expect that neutrinos with wavelength large compared to the structure size, but small compared to the inter-structure distance, can couple coherently to the total weak charge (excess) of the structure, much like the coherence we discussed in the previous section. The effects of this coherent enhancement in the neutrino cross sections has recently been investigated by \cite{Horowitz:2004yf}. In agreement with our naive expectation, their study finds that the neutrino cross sections are greatly enhanced, by as much as an order of magnitude, for neutrinos with energy  $1/r \ge E_{\nu} \ge 1/R$.  

\subsection{Nuclear liquid: Neutrino opacity at $\rho \ge 10^{14}$ g/cm$^3$}
\label{nuclearmatter}

When the density of matter exceed $10^{14}$ g/cm$^3$, matter is expected to form a homogenous liquid of nucleons and leptons. To account for the effects of strong and electromagnetic correlations between target neutrons, protons and electrons we must find ways to improve $\Pi_{\alpha\beta}$, the polarization tensor of free Fermi gas described earlier in eq. (\ref{polarization}). This involves improving the Green's functions for the particles and the associated vertex corrections that modify the current operators. In strongly coupled systems, these improvements are notoriously difficult and no exact analytic methods exist. With few exceptions, most investigations of many-body effects in the neutrino opacities have employed the mean-field theory to improve the Green's functions.

\begin{figure}
\vspace*{-0.0in}
\begin{center}
\resizebox{1.\textwidth}{!}{%
\includegraphics{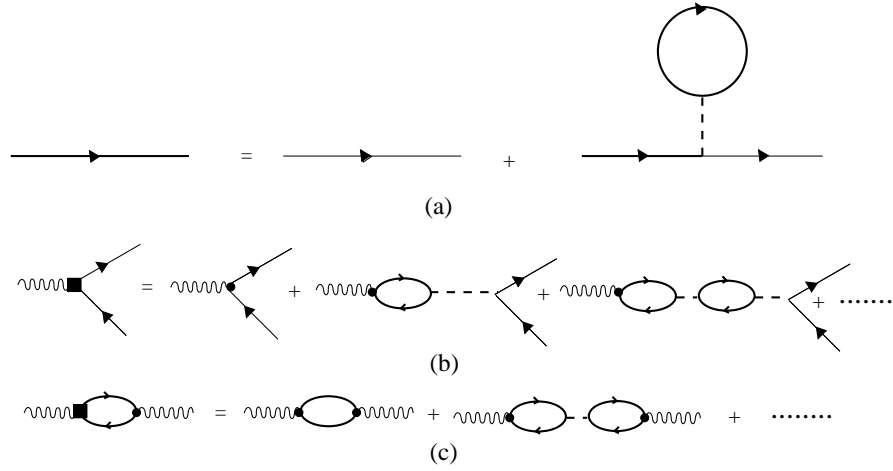}
}
\caption{Feynman diagram representing the Schwinger-Dyson equations: (a) mean field (Hartree) propogator;  (b) dressed vertex for coupling to an external current with in the RPA; and (c) RPA polarization tensor or current-current correlation function for the coupling to external currents. Dressed propagators are thick solid lines, bare propagators are thin lines, dashed lines are strong/electromagnetic interactions, wavy line is the external (weak) current, the filled circle is  the bare weak vertex, and  the filled square represents the dressed weak vertex.}
\label{schwingerdyson}
\end{center}
\end{figure}

Dressing the single particle Green's functions must be accompanied by corresponding corrections to the weak interaction vertex function.  The Green's functions in the mean field (Hartree) approximation  involves the Schwinger-Dyson summation of the diagrams shown in Fig. \ref{schwingerdyson}. The dressing of Green's functions implies the presence of an interaction ``cloud" surrounding the quasi-particle. This cloud modifies the coupling of an external probe to the quasi-particle. In \S\ref{ionplasma}, we discussed this screening of the weak charge due to Coulomb interactions between ions within RPA . Here, we consider screening due to both and strong and electromagnetic interactions.  The modification of the  coupling of an external current to the quasi-particles is shown in the lower panel of Fig. \ref{schwingerdyson}, where the solid squares represent interactions between nucleons.   

\subsubsection{Non-Relativistic RPA}

At sub-nuclear density, nucleons are non-relativistic. In the non-relativistic limit the structure  of the hadronic current greatly simplifies and the vector and axial currents can be directly related to the density and spin-density operators.
Neutrino scattering off a non-relativistic neutron liquid was first investigated by  \cite{iwamoto82}. In this case, the scattering rate can be written as
\begin{eqnarray}
\frac{d\Gamma} {d\cos\theta dE'_\nu} &=&  \frac{G_F^2}{4\pi^2} ~(1-f_\nu(E'_\nu))~{E'_\nu}^2\nonumber \\ &\times&\left({c_V}^2~(1+\cos\theta)~S(|\vec{q}|,\omega) ~+~{c_A}^2~(3-\cos\theta)~S^A(|\vec{q}|,\omega)\right)\,,\nonumber\\
\end{eqnarray}
where the dynamic structure function for density fluctuations is 
\beq
S(|\vec{q}|,\omega)=\int^{\infty}_{-\infty}dt~\exp(i\omega t)~\langle~\rho(\vec{q},t) \rho(-\vec{q},0)~\rangle \,,
\eeq
and the dynamic structure function for spin-density fluctuations is
\beq
S^A(|\vec{q}|,\omega)=\int^{\infty}_{-\infty}dt~\exp(i\omega t)~\delta_{ij}~\langle~\sigma_i(\vec{q},t) \sigma(-\vec{q},0)~\rangle \,.
\eeq
The thermal ensemble average is denoted by $\langle \cdots \rangle$. It samples the operators over all states of the many-particle system with weight $\exp(-\mathcal{H}/kT)$, where $\mathcal{H}$ is the Hamiltonian describing the neutron liquid. The formalism is easily extended to include protons (Sawyer 1975; Horowitz \& Wehrberger 1991; Burrows \& Sawyer 1998,1999).  In this case, the
differential neutrino-nucleon scattering rate from a liquid of neutrons and protons is given by,
\begin{eqnarray}
\frac{d^2 \Gamma}{d \omega \hspace{1 pt} d \cos\theta}=
(4 \pi^2)^{-1}G_F^2 E_2^2[1-f_{\nu}(E_2)]\Bigl[\bigl(1+
\cos\theta\bigr)(c_{V}^n)^2S_{nn}(q,\omega)
\nonumber\\
+\bigl(3-\cos\theta\bigr)g_{A}^2\bigl[ S^A_{pp}(q,\omega)+
S^A_{nn}(q,\omega)-2S^A_{pn}(q,\omega)  \bigr],
\label{a38}
\end{eqnarray}
where $E_2$=$E_1-\omega$.

Following Burrows \& Sawyer (1998), we consider a simple nuclear potential, with only central s-wave interactions, given by
\begin{equation}
\mathcal{V}= V_1 + V_2~\tau_1\cdot\tau_2 +
 V_3~\sigma_1 \cdot \sigma_2 + V_4~\sigma_1 \cdot \sigma_2 \tau_1 \cdot \tau_2~   \,,
\label{nuclearpotential}
\end{equation}
where $\sigma$ and $\tau$ are $2\times 2$ Pauli matrices acting in spin and iso-spin space, respectively. The structure functions, $S$ (Fermi) and $S^A$ (Gamow-Teller; axial), are elements of separate $2\times2$ symmetric matrices. For the vector dynamic structure function, $S$, we have
\begin{displaymath}
S(q,\omega)=
\begin{pmatrix}
S_{pp}(q,\omega)&S_{pn}(q,\omega)\cr S_{pn}(q,\omega)&S_{nn}(q,\omega) \, . \cr
\end{pmatrix}
\end{displaymath}

The structure function matrix is given by,

\begin{equation}
S(q,\omega)=2 {\rm Im} \Bigl[\Pi^{(0)}(q,\omega)
[1-v(q) \Pi^{(0)}(q,\omega)]^{-1}
\Bigr](1-e^{-\beta\omega})^{-1}
\label{ebetao}
\end{equation}
where

\begin{displaymath}
\Pi^{(0)}(q,\omega)=
\begin{pmatrix}
\Pi^{(0)}_p(q,\omega) &0\cr 0&\Pi^{(0)}_n (q,\omega)\cr
\end{pmatrix}
\end{displaymath}
and $\Pi^{(0)}_p$ and $\Pi^{(0)}_n$ are given by the Fermi gas polarization functions  and evaluated with
the proton and neutron chemical potentials, respectively.
The non-relativistic polarization for a Fermi gas is given by 
\begin{eqnarray}
{\rm{Im}}\Pi^{(0)}(q,\omega)&=&\frac{m^2}{2\pi q\beta}
\log\left[\frac{1+e^{-Q_+^2+\beta\mu}}{1+e^{-Q_+^2+\beta\mu-\beta\omega}}\right]\,, \\
{\rm{Re}}\Pi^{(0)}(q,\omega)&=&\mathcal{P}~\int\frac{d\omega'}{\pi}~\frac{{\rm{Im}}\Pi^{(0)}(q,\omega)}{\omega-\omega'}\, ,
\end{eqnarray}
where
\beq
Q_\pm=\left(\frac{m\beta}{2}\right)^{1/2}\left(\mp\frac{\omega}{q}+
\frac{q}{2m}\right) \, .
\eeq

The potential matrix is,

\begin{displaymath}
v=\begin{pmatrix} 
v_1+v_2+4\pi e^2 (q^2+q_{TF}^2)^{-1}&v_1-v_2\cr v_1-v_2&v_1+v_2\cr
\end{pmatrix}\,,
\label{potmatrix}
\end{displaymath}
where $v_i=\int d^3r~V_i$,  and the term containing
$q_{TF}$ is the Thomas-Fermi screened Coulomb potential ($q_{TF}^2=4 e^2\pi^{1/3}(3\bar{n}_p)^{2/3}$). In Fermi liquid theory (FLT), the underlying interaction between (quasi)particle-hole states is directly related to  thermodynamic quantities such as the susceptibilities and specific heat (see \cite{baym}  for a pedagogic introduction to FLT). These effective interactions can differ greatly from the free-space nucleon-nulceon 
potentials. In the absence of experimental determinations of the thermodynamic susceptibilities, microscopic theories of nuclear interactions are need to 
obtain these effective interactions. For the simple case of only central interactions, the susceptibilities of nuclear matter are encoded in four constants
called Landau parameters ($F_0,F_0',G_0,G_0'$ correpond to baryon density, isospin density, spin-density and spin-isospin density fluctuations, respectively).   Following Burrows \& Sawyer (1998,1999) and for simplicity, we use these Landau parameters to determine the potential matrix,  in lieu of a more developed nuclear interaction model. The parameters of the potential matrix are related to the Landau parameters by the following relations
\begin{displaymath}
\begin{matrix}
v_1+v_2=\frac{F_0+F_0'}{N_0}&v_1-v_2=\frac{F_0-F_0'}{N_0}
\cr v_3+v_4=\frac{G_0+G_0'}{N_0}&v_3-v_4=\frac{G_0-G_0'}{N_0}\cr
\end{matrix}\,  ,
\end{displaymath}
where $N_0=2m^*k_F/\pi^2$ is the density of states at the Fermi surface of nuclear matter, $m^*$ is the effective nucleon mass, and $k_F$ is the nucleon Fermi momentum. Taking  $m^*=0.75 m_n$ as our fiducial value for the effective mass, we use parameters from Backman, Brown, \& Niskanen (1985) and  Brown \& Rho (1981): $F_0= -0.28; F_0'=0.95; G_0=0;
G_0'=1.7$, obtaining,
\begin{eqnarray}
&v_1=-7.4 \times 10^{-6}\,{\rm MeV}^{-2}
\nonumber\\
&v_2=2.5\times 10^{-5}\,{\rm MeV}^{-2}
\nonumber\\
&v_3=0
\nonumber\\
&v_4=4.5\times 10^{-5}\,{\rm MeV}^{-2}.
\label{asabove}
\end{eqnarray}

For other values of the effective mass, we keep these potentials at the
same value, which is to say we assume that the Landau parameters are
proportional to  $m^*/ m$. 

The form for the Gamow--Teller matrix, $S^A(q,\omega)$, is the same as that for $S$,
except that the potential matrix is replaced by $v^A$

\begin{displaymath}
v^A=\begin{pmatrix}
v_3+v_4&v_3-v_4\cr v_3-v_4&v_3+v_4\cr
\end{pmatrix}\,.
\end{displaymath}

Taking the matrix inverses leads to the following forms for the
combinations of structure functions that appear in eq. (\ref{a38})

\begin{equation}
S_{nn}(q,\omega)=2 {\rm Im} \bigl[\Pi_n^{(0)}D_V^{-1}\bigl](1-e^{-\beta\omega})^{-1},
\label{vector}
\end{equation}
where

\begin{equation}
D_V=1-(v_1+v_ 2)\Pi_n^{(0)}-(v_1-v_2)^2\Pi_n^{(0)}\Pi_p^{(0)}{Q_V}^{-1}\, .
\label{DVterm}
\end{equation}
$Q_V$ is given by the expression:  
\begin{equation}
Q_V=1-4\pi e^2(q^2+q_{TF}^2)^{-1}\Pi_p^{(0)}-(v_1+v_2)\Pi_p^{(0)}\, .
\label{DVterm2}
\end{equation}

If, as in eq. (\ref{asabove}), we take $v_3 =0$, we obtain
the simple result for the axial--current terms,

\begin{equation}
S_A(q,\omega)=
2 {\rm Im} \Big[\frac{\Pi_p^{(0)}(q,\omega)+\Pi_n^{(0)}(q,\omega)}
{1-v_4[\Pi_p^{(0)}(q,\omega)+\Pi_n^{(0)}(q,\omega)]}\Bigr](1-e^{-\beta\omega})^{-1}\, .
\label{axial}
\end{equation}

For the Fermi term, since $ c_V^{(p)}=1/2-2\sin^2\theta_W\sim 0 $,
we drop the proton structure function in
eq. (\ref{a38}).  Furthermore, we use the potential parameters
given in eq. (\ref{asabove}), and in eq. (\ref{DVterm})
we drop the third term.  This term would have
been significant had it not been for the
Coulomb term in the denominator, an illustration of the importance
of the explicit inclusion of Coulomb forces, even for the neutron
density correlations.
Since the $v_i$s are all real,
we obtain for the structure factors used in eq. (\ref{a38}),

\begin{equation}
S_{F}(q,\omega)=2 {\rm Im} \Pi_n^{(0)}(1-e^{-\beta\omega})^{-1}{\cal C_V}^{-1},
\label{vectel}
\end{equation}
where
\begin{equation}
{\cal C_V}= (1 - v_F{\rm Re}\Pi_n^{(0)})^2 +
v_F^{2}({\rm Im}\Pi_n^{(0)})^2,
\label{vectel2}
\end{equation}
and

\begin{equation}
S_A(q,\omega)=
2 \Bigl[{\rm Im}\Pi_p^{(0)}(q,\omega)+{\rm Im}\Pi_n^{(0)}(q,\omega)\Bigr]
(1-e^{-\beta\omega})^{-1}{\cal C_A}^{-1},
\label{axel}
\end{equation}
where
\begin{equation}
{\cal C_A}={\cal C}_{{\cal A}1} + {\cal C}_{{\cal A}2}\, .
\label{axel2}
\end{equation}
${\cal C}_{{\cal A}1}$ and ${\cal C}_{{\cal A}2}$ are given by the expressions:
\begin{equation}
{\cal C}_{{\cal A}1}=
\Bigl[1-v_{GT}({\rm Re}\Pi_p^{(0)}(q,\omega)+{\rm Re}\Pi_n^{(0)}(q,\omega))\Bigr]^2
\label{axel3}
\end{equation}
and 
\begin{equation}
{\cal C}_{{\cal A}2}=
v_{GT}^{2}\Bigl[{\rm Im}\Pi_p^{(0)}(q,\omega)+{\rm Im}\Pi_n^{(0)}(q,\omega)\Bigr]^2 \, .
\label{axel4}
\end{equation}

\begin{figure}
\vspace*{-1.0in}
\begin{center}
\resizebox{.48\textwidth}{!}{%
\includegraphics{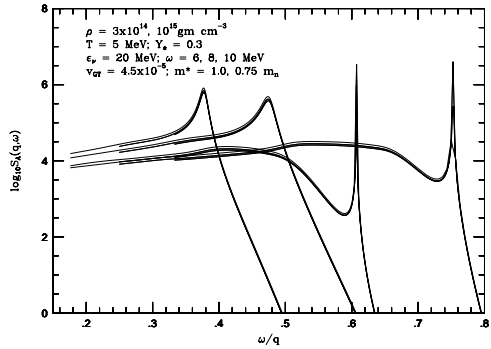}
}
\caption{Log$_{10}$ of the Gamow--Teller structure function
versus $\omega/q$ for an incident neutrino energy
of 20 MeV, energy transfers, $\omega$, of 6, 8,
and 10 MeV, two values of the effective
mass ($ m^* = [0.75m_n, 1.0 m_n]$) and two values
of the density ($\rho = 3\times10^{14}$ and $10^{15}$ g cm$^{-3}$).
A temperature of 5 MeV and a $Y_e$ of 0.3 were used,
as was the default $v_{GT}$ ($=4.5\times10^{-5}$). 
(Figure taken from Burrows and Sawyer 1998.)
}
\label{figBS2}
\end{center}
\end{figure}

The $F$ in $ S_{F}(q,\omega) $ and the $A$ in $ S_A(q,\omega) $ stand for Fermi
and Gamow--Teller (axial) and $v_F$ and $v_{GT}$ equal
$(v_1+v_ 2)$ and $v_4$, respectively, in Fermi Liquid Theory.  $S_A(q,\omega)$ in eq. (\ref{axel}) is
now the entire axial term in eq. (\ref{a38}).  ${\cal C_{V,A}}$ is the correction factor due to many-body effects for a given momentum transfer (or scattering angle) and energy transfer.  A similar procedure is employed for calculating the many-body corrections to the charged-current rates (Burrows \& Sawyer 1999).
Figure \ref{figBS2} portrays $S_A(q,\omega)$ versus $\omega/q$ for representative 
parameters and clearly indicates the resonances and \v{Cerenkov} kinematics.

\subsubsection{Multi-pair excitations and the Tensor Force}

The simple model, based on Fermi Liquid Theory and the Random Phase Approximation with central 
interactions, is rudimentary and relies on the poorly constrained determinations of the Landau
parameters. While we can expect this simple analysis to capture qualitative aspects of many-body correlations it is not suitable for quantitative predictions. It has two important shortcomings, namely the neglect of multi-pair excitations and non-central interactions such as the tensor force (pion exchange), which is known to be important in nuclear systems. The discussions in \S\ref{ionplasma} regarding the response of the ion-plasma clearly emphasized the need to incorporate finite propagation lifetimes for quasi-particles due to collisional damping. The role of collisional damping in the low-energy response is particularly well studied for the case of low-energy photon production from bremsstrahlung in many-particle systems and is called the Landau-Pomeranchuk-Migdal effect. Similar effects were first investigated in the context of the response of nuclear matter by \cite{raffelt_seckel95}. They found that the rate of spin fluctuations ($1/\tau_\mathrm{spin}$) due to nucleon-nucleon collisions in the medium is rapid compared to the typical energy transfer $\omega$ in neutrino scattering. Consequently, the axial charge is dynamically screened for small $\omega \lesssim \tau^{-1}_ \mathrm{spin}$, resulting in a suppression of the low-energy axial response. 

The redistribution of response strength in energy is a generic feature of many-particle systems arising due to the finite lifetime of quasi-particles. However, the situation in nuclear systems is unique due to the presence of a strong tensor force. This has recently been clarified by \cite{olsson_pethick}. The evolution of nucleon spin is dominated by tensor interactions, especially because the nucleon spin operator $\hat{\sigma}$ does not commute with the tensor operator in the nuclear Hamiltonian. The F-sum rule, discussed earlier in \S\ref{ionplasma}, for the spin response function does not vanish in the long-wavelength limit, since $[\mathcal{H}_{\mathrm{tensor}},\hat{\sigma}]\neq0$. Further, the relationship between the spin susceptibility  and the Landau parameters is modified due to presence of the tensor interaction (see \cite{olsson_pethick} for these revised relations). The preceding discussion indicates that interactions that do not commute with the spin operator may be especially important in determining the low-energy response.  \cite{olsson_pethick}  estimate that as much as 60\% of the low-energy axial response at long wavelength may reside in multi-particle excitations. This preliminary estimate warrants further investigation requiring  both the inclusion of the tensor force and multi-pair excitations in the axial response. 

When multi-pair excitations become important it is appropriate to work in terms of a correlated basis states rather than single particle bare neutron and proton states. The correlated basis states are expected to be close to the energy eigenstates of the system. Consequently, in this basis the residual interactions are weak.   Further, the ground and excited states in the correlated basis states contain multi-particle hole states and the quasi-particles of the correlated basis are superpositions of neutron and proton states of both spins. This is particularly relevant for nuclear matter, where pion exchange can transform both spin and isospin of the bare nucleons. Recently, \cite{cowell_pandharipande} have computed weak interaction matrix elements in the correlated basis obtained using a two-body cluster expansion. They find that spin and iso-spin correlations 
play an important role and result in quenching the weak interaction transition rates by $20-25 \%$ at low energy.

\subsubsection{Relativistic RPA}

The RPA polarization tensors that enter the neutrino scattering and absorption rates in a  relativistic frame-work have been computed  previously by Horowitz \& Wehrberger (1992) and Reddy et al. (1999). The specific form of the 
polarization tensor in relativistic random phase approximation (RRPA) follows from the structure of the bubble sum shown in Fig. \ref{schwingerdyson}.  For a one-component Fermi system, such as neutron matter, the polarization functions  
appearing in eq. (\ref{dcross1}) are replaced by 
\begin{equation}
\Pi^{RPA}_{\mu \nu} = \Pi^{\mathrm{MF}}_{\mu \nu} + \Pi^{RPA}_{\mu \alpha} D_{\alpha \beta}  \Pi^{\mathrm{MF}}_{\beta \nu}~, 
\end{equation} 
where $D_{\alpha \beta}$, is the relativistic equivalent of the potential matrix defined in \S\ref{potmatrix}. It describes the interaction between neutrons and $\Pi_{\mathrm{MF}}$ is the  polarization tensor in eq. (\ref{polar}), but with the Green's functions for neutrons computed in the mean-field approximation.  In relativistic mean-field theories (RMF), which are inspired by the  Walecka model, nucleons interact via the exchange of (fictitious) scalar, isoscalar-vector, and isovector-vector mesons (for recent review of RMF models see Serot \& Walecka 1997). The structure of $D_{\alpha \beta}$ and the mean-field Green's functions in RMF theory may be found in \cite{horowitz92}. 

The neutrino mean free paths computed in the context of these models indicate a moderate suppression of 10-20\% in the neutrino cross sections at nuclear density. Despite the strong coupling between nucleons, the suppression is modest. This is because the interactions in RMF models primarily affect the vector response, while the neutrino scattering is dominated by the axial response. The RMF model is constrained by its mean field predictions for the empirical properties of {\it spin saturated} matter. 
It is necessary to supplement the RMF model with additional spin dependent interactions such as a pion exchange which do not contribute to the ground state properties in the Hartree or mean field approximation. However, these interactions modify the RRPA response.  Model calculations in the RRPA which include spin-isospin dependent interactions (pion exchange and phenomenological short-range interactions) have been performed by \cite{reddy_1999}. The results of these  studies indicate that neutrino mean free path computed in RPA are 2-3 times larger than in the uncorrelated system (Reddy et al. 1999). This large suppression, which is similar to those encountered in the non-relativistic models, is due to repulsive forces in the spin-isospin channels. 

As mentioned earlier, RRPA is the self-consistent response of RMF models with vector and scalar interactions. It ensures thermodynamic consistency, i.e, it  satisfies the compressibility sum rules discussed in \S\ref{sumrules}.   However, the cross sections are not greatly changed by interactions. In contrast,  interactions in the spin-isospin channel do not affect the Hartree or mean-field ground state, but modify the neutrino opacity calculated in RPA. This underscores the need for detailed studies of the spin and spin-isospin response in nuclear models. To date, supernova simulations have employed equations of state (EOS) that are based on the mean-field approximation, including non-relativistic EOS's such as those  due to \cite{lattimer}. Spin and spin-isospin susceptibility of these models remain largely unconstrained by either empirical data or input from microscopic calculations of nuclear matter.

\section{$e^+e^-$ Annihilation}
\label{eplus}    

Ignoring phase-space blocking of neutrinos in the final state and taking
the relativistic limit ($m_e\rightarrow 0$), the total electron--positron
annihilation rate into neutrino--antineutrino pairs can
be written in terms of the electron and positron phase-space densities ($\f$) (Dicus 1972):
\beq
Q_{\nu_e\bar{\nu}_e}=
K_i\left(\frac{1}{m_ec^2}\right)^2\left(\frac{1}{\hbar c}\right)^6
\int\int
\f_{e^-}\f_{e^+}(\varepsilon_{e^-}^4\varepsilon_{e^+}^3+\varepsilon_{e^-}^3
\varepsilon_{e^+}^4)\,d\varepsilon_{e^-}\,d\varepsilon_{e^+}\,\,,
\label{dicusrate2}
\eeq
where $K_i=(1/18\pi^4)c\sig_o(C^{\pr 2}_V+C^{\pr 2}_A)$.  Again, $C^{\pr}_V=1/2+2\sinw$ for
electron types, $C^{\pr}_V=-1/2+2\sinw$ for $\unu$ and $\tnu$ types, and
$C^{\pr 2}_A=(1/2)^2$. Rewriting eq. (\ref{dicusrate2}) in terms of the Fermi
integral $F_n(\eta)$, we obtain:
\beq
Q_{\nu_e\bar{\nu}_e}=K_i\,(kT)\left(\frac{kT}{m_ec^2}\right)^2
\left(\frac{kT}{\hbar c}\right)^6
\left[F_4(\eta_e)F_3(-\eta_e)+F_4(-\eta_e)F_3(\eta_e)\right] \,\, ,
\eeq
where $\eta_e\equiv\mu_e/kT$ and
\beq
F_n(\eta)\equiv\int_0^\infty\,\frac{x^n}{e^{x-\eta}+1}\,dx\,\,.
\eeq
Integrating eq. (\ref{dicusrate2}), we obtain
\beq
Q_{\nu_e\bar{\nu}_e}\simeq9.7615\times 10^{24}\,\left[\frac{kT}{{\rm
MeV}}\right]^9\,f(\eta_e)\,\, {\rm ergs\, cm^{-3} s^{-1}} \, ,
\label{pairtotal}
\eeq
where
\beq
f(\eta_e)=
\frac{F_4(\eta_e)F_3(-\eta_e)+F_4(-\eta_e)F_3(\eta_e)}{2F_4(0)F_3(0)}\,\, .
\eeq
For $\nu_\mu\bar{\nu}_\mu$ and $\nu_\tau\bar{\nu}_\tau$ production
combined,
\beq
Q_{\nu_{\mu,\tau}\bar{\nu}_{\mu,\tau}}\simeq 4.1724\times
10^{24}\,\left[\frac{kT}{{\rm MeV}}\right]^9\,f(\eta_e)\,\,
{\rm ergs\, cm^{-3} s^{-1}} \, .
\label{enumurate}
\eeq

One can easily derive the spectrum of the total radiated neutrino energy ($\varepsilon_{T}$)
by inserting a delta function ($\int \delta (\varepsilon_{T}-\varepsilon_{e^-}-\varepsilon_{e^+}) d\varepsilon_{T}$)
into eq. (\ref{dicusrate2}).   Recall that the total energy of the neutrinos in the final
state is equal to the sum of the electron and positron energies in the initial state.
Integrating first over $\varepsilon_{e^+}$ to annihilate
the delta function and then over $\varepsilon_{e^-}$ to leave a function of $\varepsilon_{T}$,
one obtains:

\beq
\frac{dQ}{d\varepsilon_{T}}=
K_i\left(\frac{1}{m_ec^2}\right)^2\left(\frac{1}{\hbar c}\right)^6
\int_{0}^{\varepsilon_{T}} \varepsilon_{T} (\varepsilon_{T}-\varepsilon_{e^-})^3 \varepsilon_{e^-}^3
\f_{e^-}[\varepsilon_{e^{-}}] \f_{e^+}[\varepsilon_{T}-\varepsilon_{e^{-}}] \,d\varepsilon_{e^-}\,\, .
\label{dicusrate_et}
\eeq
The numerical evalution of eq. (\ref{dicusrate_et}) is straightforward.  The average of $\varepsilon_{T}$
is equal to:

\beq
\langle \varepsilon_{T} \rangle = \Bigl(\frac{F_4(\eta_e)}{F_3(\eta_e)} + \frac{F_4(-\eta_e)}{F_3(-\eta_e)}\Bigr)T\, ,
\label{averageet}
\eeq
which near $\eta_e \sim 0$ is $\sim 8T$ and for $\eta_e >> 1$ is $\sim 4T(1+\eta_e/5)$.

However, while the total energy loss rate (eq. \ref{pairtotal}) and the spectrum of $\varepsilon_{T}$
pose no great mathematical problems, the production spectrum of an individual neutrino is not so easily reduced to a
simple integral or to an analytic expression.  This is due primarily to the awkward
integration of the angular phase space terms, while subject to the momentum conservation
delta function, and to the explicit dependence of the matrix elements
on the electron/neutrino angles.
From \cite{dicus72}, averaging over initial states and summing over final states, the
matrix element for the $e^+e^-\to\nu\bar{\nu}$ process in the $m_e=0$ limit is:
\beq
\frac{1}{4}\sum_s|{\cal {M}}|^2=16G^2[(C^{\pr}_V+C^{\pr}_A)^2{\bf p \cdot
q}_{\bar{\nu}}\,{\bf p}\pr\cdot{\bf q}_\nu+(C^{\pr}_V-C^{\pr}_A)^2{\bf p\cdot
q}_\nu\,{\bf p}\pr\cdot{\bf q}_{\bar{\nu}}]\,\, ,
\label{avematrix}
\eeq
where $p$ and $p\pr$ are the four-momenta of the electron and positron,
respectively, and $q_\nu$ and $q_{\bar{\nu}}$ are the four-momenta of the
neutrino and antineutrino, respectively.  Using the formalism of
\cite{bruenn_1985} and
Fermi's Golden rule, expanding the production kernel in the traditional truncated
Legendre series, performing the trivial angular integrals, taking the
non--trivial angular integrals from \cite{bruenn_1985}, and ignoring final--state
neutrino blocking, we obtain for the single--neutrino
source spectrum due to $e^+e^-$ annihilation:
\beq
\frac{dQ}{d\varepsilon_\nu}=\,\frac{8\pi^2}{(2\pi \hbar
c)^6}\,\varepsilon_\nu^3\,\int_0^\infty\,d\varepsilon_{\bar{\nu}}\,
\varepsilon_{\bar{\nu}}^2\,\Phi_0^p(\varepsilon_\nu,\varepsilon_{\bar{\nu}})\,\, ,
\label{qspectrum}
\eeq
where
\beq
\Phi_0^p(\varepsilon_\nu,\varepsilon_{\bar{\nu}})=\frac{G^2}{\pi}
\int_0^{\varepsilon_\nu+\varepsilon_{\bar{\nu}}}d\varepsilon_{e^-}{\cal
{F}}_{e^-}[\varepsilon_{e^-}]{\cal
{F}}_{e^+}[\varepsilon_\nu+\varepsilon_{\bar{\nu}}-\varepsilon_{e^-}]\,
H_0(\varepsilon_\nu,\varepsilon_{\bar{\nu},}\varepsilon_{e^-})\,\,,
\label{phi0}
\eeq
and
\beq
H_0(\varepsilon_\nu,\varepsilon_{\bar{\nu}},\varepsilon_{e^-})=(C^{\pr}_V+C^{\pr}_A)^2\,
J_0^I(\varepsilon_\nu,\varepsilon_{\bar{\nu}},\varepsilon_{e^-})+(C^{\pr}_V-C^{\pr}_A)^2
\,J_0^{II}(\varepsilon_\nu,\varepsilon_{\bar{\nu}},\varepsilon_{e^-})\,\, .
\label{polynomials}
\eeq
The $J_0$s in eq. (\ref{polynomials}) come from the more obdurate angular
integrals required by the dot products in eq. (\ref{avematrix}) and the momentum
delta function and have the symmetry:
\beq
J_0^I(\varepsilon_\nu,\varepsilon_{\bar{\nu}},\varepsilon_{e^-})=J_0^{II}
(\varepsilon_{\bar{\nu}},\varepsilon_\nu,\varepsilon_{e^-})\,\, .
\eeq
From eqs. (\ref{qspectrum}) and (\ref{polynomials}), we see that the differences
between the spectra of the $\nu_e$ and $\nu_{\mu}$ neutrinos
flow solely from their correspondingly different values of $(C^{\pr}_V+C^{\pr}_A)^2$ and $(C^{\pr}_V-C^{\pr}_A)^2$.
One can use 4--point Gauss--Legendre integration to calculate eq. (\ref{phi0}) and 16--point
Gauss--Laguerre integration to calculate eq. (\ref{qspectrum}).

At small $\eta_e$, the $e^+ e^-$ annihilation spectra and total energy loss rates for
the $\nu_e$ and $\bar{\nu}_e$ neutrinos are similar, as are the
average emitted  $\nu_e$ and $\bar{\nu}_e$ neutrino energies.  However,
as $\eta_e$ increases, both the total energy radiated in $\bar{\nu}_e$
neutrinos and the average $\bar{\nu}_e$ energy start to lag the corresponding quantities for the $\nu_e$ neutrinos.
This is true despite the fact that the total number of $\nu_e$ and $\bar{\nu}_e$ neutrinos radiated
is the same. If final--state blocking is
ignored, $\langle \varepsilon_i \rangle/T$ is a function of $\eta_e$ alone, becoming linear with $\eta_e$
at high $\eta_e$ and one half of eq. (\ref{averageet}) ($\sim$4.0) at low $\eta_e$.
Note also that $\langle \varepsilon_{\nu_{\mu}} \rangle/T$ and $\langle \varepsilon_{\bar{\nu}_{\mu}} \rangle/T$
are closer to one another than are $\langle \varepsilon_{\nu_e} \rangle/T$ and $\langle \varepsilon_{\bar{\nu}_e} \rangle/T$.
The individual production spectra vary in peak strength, in peak energy, and
in low--energy shape, but they are quite similar on the high--energy tail.  Due to the parity--violating
matrix element for the $e^+ e^- \rightarrow \nu \bar{\nu}$ process and the fact that $\eta_e$ is positive,
the antineutrino spectra of all species are softer than the neutrino spectra.
The pair sums of the integrals under these curves are given by eqs. (\ref{pairtotal}) and (\ref{enumurate}).
For $\eta_e = 0$, 50\% of the pair energy emission of electron types is in $\bar{\nu}_e$ neutrinos, but
at $\eta_e = 10$ only 42\% of this total energy is in $\bar{\nu}_e$ neutrinos.  However, at $\eta_e = 10$,
the $\bar{\nu}_{\mu}$ neutrinos still constitute 48.5\% of the $\nu_{\mu} / \bar{\nu}_{\mu}$ pair emission.
These differences reflect differences in the corresponding coupling constants $C^{\pr}_V$ and $C^{\pr}_A$.

\section{$\nu_i\anu_i$ Annihilation}
\label{paira}  

In the limit of high temperatures and ignoring electron phase space
blocking, the  $\nu_i\anu_i$ annihilation rate into $e^+e^-$ pairs can be
written (Janka 1991):
\beq
Q_{\nu_i\anu_i}=
4K_i\pi^4\left(\frac{1}{m_ec^2}\right)\,\left(\frac{4\pi}{c}\right)^2
\int\int\,\Phi^{\prime}\,J_{\nu_i}J_{\bar{\nu}_i}(\vep_{\nu_i}+\vep_{\bar{\nu}_i})
\,d\vep_{\nu_i}\,d\vep_{\bar{\nu}_i}\,\, ,
\label{jankarate1}
\eeq
where $J_\nu$ is the zeroth moment of the radiation field,
$\epnu$ is the neutrino energy, $K_i$ is defined as before
({\it i.e.,} $K_i=(1/18\pi^4)c\sig_o(C^{\pr 2}_V+C^{\pr 2}_A)$), and
\beq
\Phi^{\prime}\left(\avemu,\aveamu,p_{\nu_i},p_{\bar{\nu}_i}\right)=
\frac{3}{4}
\left[1-2\avemu\aveamu+p_{\nu_i}p_{\bar{\nu}_i}+
\frac{1}{2}(1-p_{{\nu}_i})(1-p_{\bar{\nu}_i})\right] ,
\eeq
where the flux factor $\avemu$ = $H_\nu/J_\nu $ and the Eddington factor
$p_\nu=\avemut=P_\nu/J_\nu $. Eq.(\ref{jankarate1}) can be rewritten in
terms of the invariant distribution functions $\f_\nu $:
\beq
Q_{\nu_i\anu_i}=
K_i\,\left(\frac{1}{m_ec^2}\right)^2\left(\frac{1}{\hbar c}\right)^6
\int\int\,\Phi^{\prime}\,
\f_{\nu_i}\f_{\bar{\nu}_i}
(\varepsilon_{\nu_i}^4\varepsilon_{\bar{\nu}_i}^3+\varepsilon_{\nu_i}^3\varepsilon_{\bar{\nu}_i}^4)
\,d\varepsilon_{\nu_i}\,d\varepsilon_{\bar{\nu}_i}\, .
\label{jankarate2}
\eeq

Note that when the radiation field is isotropic ($\Phi^{\prime}=1$) and when $\eta_e=0$
the total rate for $e^+e^-$ annihilation given in eq. (\ref{dicusrate2})
equals that for $\nu_i\bar{\nu}_i$ annihilation given in eq. (\ref{jankarate2}), as expected.  
Buras et al. (2003a) have addressed the
related and interesting process of $\nu_i\bar{\nu}_i \rightarrow \nu_j\bar{\nu}_j$.
We refer to that paper for a discussion of the relevance and rates of this process.

\section{Nucleon--Nucleon Bremsstrahlung}
\label{bremsst}

A production process for neutrino/anti-neutrino pairs that has recently received 
attention in the supernova context is neutral-current
nucleon--nucleon bremsstrahlung ($n_1 + n_2 \righta n_3 + n_4 + \nu\bar{\nu}$).
It importance in the cooling of old neutron stars, for which the nucleons are quite
degenerate, has been recognized for years (Flowers 1975), but only
in the last few years has it been studied for its potential importance 
in the quasi-degenerate to non-degenerate atmospheres of protoneutron stars
and supernovae (Suzuki 1993; Hannestad \& Raffelt 1998; Burrows et al. 2000; 
Thompson, Burrows, \& Horvath 2000).  Neutron--neutron, proton--proton, and neutron--proton
bremsstrahlung are all important, with the latter the most important for symmetric matter.  As a source of $\nu_e$ and
$\bar{\nu}_e$ neutrinos, nucleon--nucleon bremsstrahlung can not compete
with the charged--current capture processes.
However, for a range of temperatures and densities realized in
supernova cores, it may compete with $e^+e^-$ annihilation as a source
for $\nu_\mu$, $\bar{\nu}_\mu$, $\nu_\tau$, and $\bar{\nu}_\tau$ neutrinos (``$\nu_\mu$''s).
The major obstacles to obtaining accurate estimates of the emissivity of this process are our poor knowledge
of the nucleon--nucleon potential, of the degree of suitability of the Born Approximation, and
of the magnitude of many--body effects (Hannestad \& Rafflet 1998; Raffelt \& Seckel 1998; 
Brinkmann \& Turner 1988).
Since the nucleons in protoneutron star atmospheres are not degenerate,
we present here a calculation of the total and differential
emissivities of this process in that limit and assume a
one-pion exchange (OPE) potential model to calculate the
nuclear matrix element.  For the corresponding calculation for arbitrary nucleon degeneracy,
the reader is referred to \cite{thompson}.  
The formalism we employ has been heavily influenced by those of  
\cite{brinkmann} and \cite{han.raff},
to which the reader is referred for details and further explanations.

Our focus is on obtaining
a useful single--neutrino final--state emission (source) spectrum, as well as a final--state pair energy spectrum
and the total emission rate.  For this, we start with Fermi's Golden Rule for the total rate per neutrino species:
\beqa
Q_{nb}=(2\pi)^4\int \Bigl[\prod_{i=1}^4 \frac{d^3\vec{p}_{i}}{(2\pi)^3}\Bigr]
\frac{d^3\vec{q}_\nu}{(2\pi)^3 2\omega_\nu}
\frac{d^3\vec{q}_{\bar{\nu}}}{(2\pi)^3 2\omega_{\bar{\nu}}}\,
\omega\, \sum_{s}{|{\cal{M}}|}^2 \delta^4({\bf P})\, \Xi_{brems},
\nonumber 
\eeqa
where
\beqa
\Xi_{brems}=\f_1\f_2(1-\f_3)(1-\f_4),
\label{bremfermi}
\eeqa
$\delta^4({\bf P})$ is four--momentum conservation delta function,
$\omega$ is the energy of the final--state neutrino pair,
($\omega_\nu$,$\vec{q}_\nu$) and ($\omega_{\bar{\nu}}$,$\vec{q}_{\bar{\nu}}$)
are the energy and momentum of the neutrino and anti--neutrino, respectively,
and $\vec{p}_{i}$ is the momentum of nucleon $i$.  Final--state
neutrino and anti--neutrino blocking have been dropped.

The necessary ingredients for the integration of eq. (\ref{bremfermi})
are the matrix element for the interaction and a workable procedure for handling
the phase-space terms, constrained by the conservation laws.   We follow 
\cite{brinkmann} for both of these elements. In particular, we assume for the
$n + n \righta n + n + \nu\bar{\nu}$ process that the
matrix element is:

\beqa
\sum_{s}{|{\cal{M}}|}^2 = \frac{64}{4} G_F^2(f/m_\pi)^4 g_A^2 \Bigl[ (\frac{k^2}{k^2+m_{\pi}^2})^2 + \dots \Bigr ]
\frac{\omega_{\nu} \omega_{\bar{\nu}}}{\omega^2}
\nonumber \\
=A\frac{\omega_{\nu} \omega_{\bar{\nu}}}{\omega^2} \, ,
\label{matrixbrem}
\eeqa
where the $4$ in the denominator accounts for the spin average for identical nucleons, $G_F$ is the
weak coupling constant, $f$ ($\sim1.0$) is the pion--nucleon coupling constant, $g_A$ is the axial--vector
coupling constant, the term in brackets is from the OPE propagator plus exchange and cross terms, $k$ is the nucleon
momentum transfer, and $m_\pi$ is the pion mass.   In eq. (\ref{matrixbrem}), we have dropped $\vec{q}_\nu\cdot\vec{k}$
terms from the weak part of the total matrix element.  To further simplify the calculation, we set the
``propagator'' term equal to a constant $\zeta$, a number of order unity, and absorb into
$\zeta$ all interaction ambiguities.  

Recently, \cite{hanhart} have addressed these
momentum terms in the context of axion emission and $\nu_\mu\bar{\nu}_\mu$
production in supernovae.  In an effort to make contact with the approximation
to the matrix element we present here, they plot $\zeta$ as a function
of average relative thermal nucleon momentum ($\bar{p}$; Phillips, private communication).
The function peaks for $\zeta(\bar{p})$ between $150-200$ MeV at $\zeta\simeq0.47$.
At $\bar{p}=50$ MeV  $\zeta\simeq0.08$ and at $\bar{p}=500$ MeV $\zeta\simeq0.27$.
We are most interested in the region around the $\nu_\mu$ neutrinospheres,
where the emergent spectrum might be most affected by nucleon-nucleon bremsstrahlung.
Mass densities and temperatures in this region might be $10^{12}-10^{13}$ g cm$^{-3}$ and
$5-10$ MeV, respectively.  We estimate $\bar{p}$ in this regime to be $\sim175$ MeV
and take $\zeta=0.5$ for all thermodynamical points. The constant $A$ in eq. (\ref{matrixbrem}) remains.

Inserting a $\int \delta(\omega - \omega_{\nu} - \omega_{\bar{\nu}})d\omega$ by the neutrino phase space terms
times $\omega \omega_{\nu} \omega_{\bar{\nu}}/{\omega^2}$ and integrating over $\omega_{\bar{\nu}}$ yields:

\beq
\int \omega \frac{\omega_{\nu} \omega_{\bar{\nu}}}{\omega^2} \frac{d^3\vec{q}_\nu}{(2\pi)^3
2\omega_\nu}\frac{d^3\vec{q}_{\bar{\nu}}}{(2\pi)^3 2\omega_{\bar{\nu}}}\righta\frac{1}{(2\pi)^4}
\int_{0}^{\infty} \int_{0}^{\omega} \frac{\omega_{\nu}^2 (\omega - \omega_{\nu})^2}{\omega} d\omega_{\nu} d\omega \,  ,
\label{deltaneut}
\eeq
where again $\omega$ equals ($\omega_{\nu} + \omega_{\bar{\nu}}$).  If we integrate
over $\omega_{\nu}$, we can derive the $\omega$ spectrum.  A further integration over $\omega$
will result in the total volumetric energy emission rate.  If we delay such an integration, after
the nucleon phase space sector has been reduced to a function of $\omega$ and if we
multiply eq. (\ref{bremfermi}) and/or eq. (\ref{deltaneut}) by $\omega_{\nu}/\omega$,  an integration
over $\omega$ from $\omega_{\nu}$ to infinity will leave the emission spectrum for the single final--state
neutrino.  This is of central use in multi--energy group transport calculations and
with this differential emissivity and Kirchhoff's Law (\S\ref{stimabs}) we can derive an absorptive opacity.

Whatever our final goal, we need to reduce the nucleon phase space integrals and to do this we use the
coordinates and approach of \cite{brinkmann}.  We define new momenta: $p_+ = (p_1 + p_2)/2$, $p_- = (p_1 - p_2)/2$,
$p_{3c} = p_3 - p_+$, and $p_{4c} = p_4 - p_+$, where nucleons $1$ and $2$ are in the initial state.  Useful direction cosines
are $\gamma_1 = p_+ \cdot p_-/|p_+||p_-|$ and $\gamma_c = p_+ \cdot p_{3c}/|p_+||p_{3c}|$.
Defining $u_i = p_i^2/2mT$ and using energy and momentum conservation, we can show that:
\beqa
d^3p_1d^3p_2 &=& 8d^3p_+d^3p_-
\nonumber \\
\omega &=& 2T(u_- - u_{3c})
\nonumber \\
u_{1,2} &=& u_+ + u_- \pm 2(u_+u_-)^{1/2}\gamma_1
\nonumber \\
u_{3,4} &=& u_{+} + u_{3c} \pm 2(u_+u_{3c})^{1/2}\gamma_c \, .
\label{upm}
\eeqa

In the non--degenerate limit, the $\f_1\f_2(1-\f_3)(1-\f_4)$ term reduces to $e^{2y} e^{-2(u_+ + u_-)}$,
where $y$ is the nucleon degeneracy factor.  Using eq. (\ref{upm}), we see that the quantity $(u_+ + u_-)$ is independent
of both $\gamma_1$ and $\gamma_c$.  This
is a great simplification and makes the angle integrations trivial.
Annihilating $d^3p_4$ with the momentum delta function in eq. (\ref{bremfermi}), noting that $p_i^2dp = \frac{(2mT)^{3/2}}{2}u_i^{1/2}du_i$,
pairing the remaining energy delta function with $u_-$, and integrating $u_+$ from $0$ to $\infty$, we obtain:
\beq
d Q_{nb} = \frac{Am^{4.5}}{2^8\times3\times5 \pi^{8.5}} T^{7.5} e^{2y} e^{-\omega/T}
(\omega/T)^4 \Bigl[ \int_0^{\infty} e^{-x}(x^2 + x\omega/T)^{1/2} dx\Bigr] d\omega \, .
\label{ezz4}
\eeq
The variable $x$ over which we are integrating in eq. (\ref{ezz4}) is equal to $2u_{3c}$.  That integral is analytic and
yields:
\beq
\int_0^{\infty} e^{-x}(x^2 + x\omega/T)^{1/2} dx = \eta e^{\eta}K_1(\eta)\, ,
\label{kintegral}
\eeq
where $K_1$ is the standard modified Bessel function of imaginary argument, related to the Hankel functions, and
$\eta = \omega/2T$.  Hence, the $\omega$ spectrum is given by:
\beq
\frac{d Q_{nb}}{d\omega} \propto e^{-\omega/2T} \omega^5 K_1(\omega/2T) \, .
\label{omegaspect}
\eeq

It can easily be shown that $\langle \omega \rangle = 4.364 T$.
Integrating eq. (\ref{ezz4}) over $\omega$ and using the thermodynamic identity in the non--degenerate limit:
\beq
e^y = \Bigl(\frac{2\pi}{mT}\Bigr)^{3/2} n_n/2 \, ,
\eeq
where $n_n$ is the density of neutrons (in this case), we derive for the
total neutron--neutron bremsstrahlung emissivity of a single neutrino pair:
\beq
Q_{nb} = 1.04\times10^{30}
\zeta(X_n \rho_{14})^2 (\frac{T}{{\rm MeV}})^{5.5} \, {\rm ergs\, cm^{-3}\, s^{-1}} \, ,
\label{bremssr}
\eeq
where $\rho_{14}$ is the mass density in units of $10^{14}$ gm cm$^{-3}$ and
$X_n$ is the neutron mass fraction.  Interestingly,  this is
within 30\% of the result in \cite{suzuki_93}, even though 
he has substituted, without much justification, $(1+\omega/2T)$ for
the integral in eq. (\ref{ezz4}) ($[1+(\pi\eta/2)^{1/2}]$ is a better
approximation). The proton-proton and neutron-proton processes can be handled similarly and the total
bremsstrahlung rate is then obtained by substituting $X_n^2 + X_p^2 + \frac{28}{3} X_n X_p$ for
$X_n^2$ in eq. (\ref{bremssr}) (Brinkmann \& Turner 1988).
At $X_n = 0.7$, $X_p = 0.3$, $\rho = 10^{12}$ gm cm$^{-3}$, and T = 10 MeV, and taking the
ratio of augmented eq. (\ref{bremssr}) to eq. (\ref{enumurate}),
we obtain the promising ratio of $\sim 5\zeta$.
Setting the correction factor $\zeta$ equal to $\sim0.5$ (Hanhart, Phillips, \& Reddy 2001), we find
that near and just deeper than the $\nu_\mu$ neutrinosphere, bremsstrahlung is larger than classical
pair production.

If in eq. (\ref{deltaneut}) we do not integrate over $\omega_\nu$, but at the
end of the calculation we integrate over $\omega$ from $\omega_\nu$ to $\infty$,
after some manipulation we obtain the single neutrino emissivity spectrum:
\beqa
\frac{d Q_{nb}^{\prime}}{d\omega_{\nu}} =
2C \Bigl(\frac{Q_{nb}}{T^4}\Bigr)
\omega_{\nu}^3 \int^{\infty}_{\eta_\nu}  \frac{e^{-\eta}}{\eta} K_1(\eta) (\eta - {\eta_\nu})^2 d\eta
\eeqa
\beqa
= 2C \Bigl(\frac{Q_{nb}}{T^4}\Bigr)
\omega_{\nu}^3 \int^{\infty}_{1} \frac{e^{-2\eta_{\nu}\xi}}{\xi^3} (\xi^2-\xi)^{1/2} d\xi \, ,
\label{spectrum}
\eeqa
where $\eta_{\nu} = \omega_\nu/2T$, $C$ is the normalization constant equal
to $\frac{3\times5\times7\times11}{2^{11}}$ ($\cong 0.564$), and for the second expression we have used the
integral representation of $K_1(\eta)$ and reversed the order of integration.  In eq. (\ref{spectrum}),
$Q_{nb}$ is the emissivity for the pair.

Eq.(\ref{spectrum}) is the approximate neutrino emission spectrum  due to nucleon--nucleon bremsstrahlung.
A useful fit to eq. (\ref{spectrum}), good to better than 3\% over the full range of important values of $\eta_{\nu}$, is:
\beq
\frac{d Q_{nb}^{\prime}}{d\omega_{\nu}} \cong
\frac{0.234 Q_{nb}}{T} \Bigl(\frac{\omega_\nu}{T}\Bigr)^{2.4} e^{-1.1 \omega_{\nu}/T}\, .
\eeq
Thompson, Burrows, \& Horvath (2000) should be consulted for a detailed discussion 
of nucleon-nucleon bremsstrahlung for arbitrary nucleon 
degeneracy.

\section{Conclusion}
 
The processes that have been described in this paper are essential elements  
of the neutrino-driven supernova explosion mechanism.  Coupling these with radiation-hydrodynamics
codes, an equation of state, beta-decay and electron capture microphysics, and
nuclear rates, one explores the viability of various scenarios for the explosion
of the cores of massive stars (Liebend\"orfer et al. 2001ab; 
Rampp \& Janka 2000,2002).  Recently, Thompson, Burrows, \& Pinto (2003) have 
incorporated this neutrino microphysics into simulations of 1D (spherical) core collapse and
have investigated the effects on the dynamics, luminosities, and emergent 
spectra of weak magnetism/recoil, nucleon-nucleon bremsstrahlung, inelastic
neutrino-electron scattering, and a host of the cross section corrections described
above.  
%
%
The character of the spectra reflect the opacities and sources.  In particular, the
energy hardness hierarchy from $\nu_e$ (softer) to $\nu_{\mu}$ (harder) neutrinos
is clearly manifest, as is the distinction between the $\nu_e$ pre-breakout
and post-breakout spectra.

To date, none of the detailed 1D simulations that have been performed 
explodes and it may be that multi-dimensional effects play a pivotal role in the explosion mechanism 
(Herant et al.~1994; Burrows, Hayes, \& Fryxell 1995; Janka \& M\"uller 1996; Fryer et al. 1999; 
Fryer \& Warren 2002).
Be that as it may, an understanding of neutrino-matter interactions remains
central to unraveling one of the key mysteries of the nuclear universe in which
we live.  

\acknowledgments

We would like to thank Joe Carlson, Christoph Hanhart, 
Chuck Horowitz, Jorge Horvath, Jim Lattimer,  Daniel Phillips,  
Madappa Prakash, and Ray Sawyer,  for fruitful discussions 
and/or collaboration on some of the more thorny 
aspects of neutrino-matter interactions.  
Support for this work was provided in part by
the Scientific Discovery through Advanced Computing (SciDAC) program
of the DOE, grant number DE-FC02-01ER41184.  T.A.T. is supported 
by NASA through Hubble Fellowship
grant \#HST-HF-01157.01-A awarded by the Space Telescope Science
Institute, which is operated by the Association of Universities for
Research in Astronomy, Inc., for NASA, under contract NAS 5-26555.

\newpage

\end{document}